\documentclass[reprint,superscriptaddress,prx]{revtex4-1} 
\usepackage[toc,page]{appendix}
\usepackage[british]{babel}
\usepackage[utf8]{inputenc}
\usepackage{amsmath,amssymb}
\usepackage{graphicx}
\usepackage{ae}
\usepackage{esdiff}
\usepackage{braket}
\usepackage{bbm}
\usepackage{simplewick}
\usepackage{float}
\usepackage{braket}
\usepackage{simplewick}
\usepackage{color}
\usepackage{framed}
\usepackage[caption=false]{subfig}
\usepackage[pdftex,colorlinks=true,linkcolor=blue,citecolor=blue,urlcolor=black]{hyperref}
\usepackage{mathtools}
\usepackage{enumitem}
\usepackage{ragged2e}
\usepackage{floatflt}
\usepackage{import}
\usepackage{bm}
\usepackage{titlesec}
\usepackage{multirow}
\usepackage{etoolbox}
\usepackage{appendix}

\makeatletter
\patchcmd{\ttlh@hang}{\parindent\z@}{\parindent\z@\leavevmode}{}{}
\patchcmd{\ttlh@hang}{\noindent}{}{}{}
\makeatother

\begin{document}

\newcommand{\norm}[1]{\left\lVert#1\right\rVert}

\renewcommand\thesubsection{\arabic{subsection}.}
\renewcommand{\arraystretch}{2}

\titleformat{\section}
  {\normalfont\bfseries}{\thesection.}{1em}{\centering}

\titleformat{\subsection}
  {\normalfont\itshape}{\thesubsection}{1em}{\centering}

\def \k{{\textbf{k}}}
\def \p{{\textbf{p}}}
\def \q{{\textbf{q}}}
\def \x{{\textbf{x}}}
\def \A{{\textbf{{A}}}}
\def \a{{\textbf{{a}}}}
\def \b{{\textbf{{b}}}}
\def \c{{\textbf{{c}}}}
\def \z{{\textbf{{z}}}}
\def \0{{\textbf{{0}}}}
\def \intk{{\int_\textbf{k}}}
\def \dl{\frac{\partial}{\partial l}}

\def\BigColSep{\setlength{\arraycolsep}{50pt}}

\title{Deconfined quantum criticality in SU(3) antiferromagnets on the triangular lattice}
\author{Dimitri Pimenov}
\affiliation{Physics Department, Arnold Sommerfeld Center for Theoretical Physics and Center for NanoScience, Ludwig-Maximilians-University Munich, 80333 Munich, Germany}
\author{Matthias Punk}
\affiliation{Physics Department, Arnold Sommerfeld Center for Theoretical Physics and Center for NanoScience, Ludwig-Maximilians-University Munich, 80333 Munich, Germany}

\begin{abstract}
We propose field theories for a deconfined quantum critical point in $SU(3)$ antiferromagnets on the triangular lattice. In particular we consider the continuous transition between a magnetic, three-sublattice color-ordered phase and a trimerized $SU(3)$ singlet phase.
Starting from the magnetically ordered state we derive a critical theory in terms of fractional bosonic degrees of freedom, in close analogy to the well-developed description of the $SU(2)$ N\'eel - valence bond solid (VBS) transition on the square lattice. Our critical theory consists of three coupled $CP^2$ models and we study its fixed point structure using a functional renormalization group approach in a suitable large $N$ limit. We find a stable critical fixed point and estimate its critical exponents, thereby providing an example of deconfined criticality beyond the universality class of the $CP^N$ model. In addition we present a complementary route towards the critical field theory by studying topological defects of the trimerized $SU(3)$ singlet phase. 
\end{abstract}
\maketitle

\section{Introduction}

Deconfined criticality is a concept that has emerged in recent years to describe quantum phase transitions beyond the Landau-Ginzburg paradigm \cite{senthil2004quantum,senthil2004deconfined,
sachdev2008quantum}. Its basic idea is that a continuous quantum phase transition between two different symmetry broken phases is generically possible, if it is driven by the proliferation of topological defects which carry quantum numbers related to the order parameter of the other phase. Disordering one phase by condensing topological defects thus automatically leads to the appearance of the other order parameter. Such continuous transitions do not arise in the Landau-Ginzburg framework, where transitions between two different symmetry broken phases are generically of first order.

The prime example for deconfined criticality is the transition between a magnetically ordered N\'eel state and a valence bond solid (VBS) in $SU(2)$ antiferromagnets on the square lattice \cite{senthil2004quantum,senthil2004deconfined,
sachdev2008quantum,read1989valence,read1990spin}. The N\'eel state spontaneously breaks the spin rotation symmetry, whereas the VBS state spontaneously breaks lattice symmetries; the transition can be driven e.g.~by changing the relative strength of nearest-neighbor exchange and ring-exchange terms in generalized Heisenberg models \cite{sandvik2007evidence}. On both sides of the transition elementary excitations (either spin waves in the N\'eel phase, or triplet excitations in the VBS phase) carry spin $S=1$, while the relevant low energy degrees of freedom at the critical point are fractionalized (or deconfined) $S=1/2$ spinon excitations, which are strongly coupled to an emergent $U(1)$ gauge field.

Early indications for the existence of such critical points came about by realizing that the nonlinear sigma model, describing low energy fluctuations of the N\'eel state in $(2+1)$-dimensions, breaks rotation symmetries in the paramagnetic phase \cite{read1989valence,read1990spin}. This is due to the fact that Berry phase terms play a crucial role if singular configurations of the N\'eel order parameter field become important \cite{Haldane1988}. These singular configurations are topological defects known as "hedgehogs", which start to proliferate and condense at the transition out of the N\'eel state and can be viewed as magnetic monopoles of the dual gauge theory. Since the monopole operator transforms nontrivially under lattice symmetries, the proliferation of monopoles automatically gives rise to VBS order. In hindsight, the fact that the paramagnetic state has to break lattice symmetries comes as no surprise. This is because unique paramagnetic ground states with an energy gap do not exist in models with one spin $S=1/2$ per unit cell in two dimensional systems, due to the Lieb-Schultz-Mattis-Hastings-Oshikawa theorem \cite{lieb1961two,hastings2004lieb,oshikawa2000commensurability}.

Alternatively one can understand the deconfined critical point by approaching it from the VBS phase. In this case the N\'eel state can be viewed as condensate of vortices on the VBS side, which carry spin $S=1/2$ and are the electric charges of the dual gauge theory \cite{levin2004deconfined}. A crucial point in the theory of deconfined criticality is that the density of monopoles vanishes at the critical point and the vortices are thus deconfined. 

An action for the critical theory can be readily derived by fractionalizing the Neel order parameter $\hat{N}$ in terms of bosonic spinor variables $z_\alpha$ ($\alpha \in \{1,2\}$) as \cite{motrunich2004emergent,senthil2004quantum,
senthil2004deconfined,sachdev2011quantum}
\begin{align}
\label{zmappingsu2}
\hat{N} = \bar{z}_\alpha \boldsymbol{\sigma}_{\alpha \beta} z_\beta  \ .
\end{align}
where $\boldsymbol{\sigma}$ is the vector of Pauli matrices. Physically, the fields $z_\alpha$ can be identified with XY-type vortices in the VBS phase which carry spin-1/2 and thus transform as spinors under $SU(2)$. One can now construct the most general action by an expansion in powers and gradients of $z_\alpha$ that are allowed by symmetries. In addition, the $U(1)$ gauge redundancy $z_\alpha \to e^{i \phi} z_\alpha$ in the mapping (\ref{zmappingsu2}) has to be incorporated by introducing a gauge field $\mathcal{A}$. The resulting theory is the celebrated euclidean $CP^1$ model in $2+1$ dimensions
\begin{align}
\label{cp1model}
S_{\text{CP}^1} = \int\! d^3 x \bigg[ &\sum_{\alpha,\mu}|(\partial_\mu -i \mathcal{A}_\mu)z_{\alpha}|^2 + m z_\alpha \bar{z}_\alpha  \nonumber \\  & + \rho (z_\alpha \bar{z}_\alpha)^2 + \frac{1}{4e^2} \mathcal{F}^2_{\mu\nu}\ \bigg]\ ,
\end{align} 
where the last term is the usual Maxwell term for the gauge field. Note that the gauge field in Eq.~\eqref{cp1model} is not compact, i.e.~monopoles are irrelevant and the gauge theory is deconfined.
The theory in Eq.~\eqref{cp1model} is strongly coupled and reliable results only exist in the large $N$ limit of generalized $CP^{N-1}$ models, where the field $z_\alpha$ has $N$ components. Nevertheless, extensive theoretical work indicates the presence of a stable critical fixed point at the relevant value $N=2$, suggesting that a second order phase transition indeed exists \cite{halperin1974first, PhysRevLett.47.1556,motrunich2004emergent, bergerhoff1996phase, bartosch2013corrections,huh2013vector}. By contrast, numerical studies of the $SU(2)$ Heisenberg model with ring-exchange terms have not reached a consensus yet. While some early works claimed evidence for deconfined criticality \cite{sandvik2007evidence, MelkoKaul2008}, later strong corrections to scaling were found \cite{Sandvik2010, Banjeree2010,shao2016quantum}, while other works claim that the transition is weakly first order \cite{Kuklov2008, Svistunov2013PRL}. The situation is much clearer for particular $SU(N)$ generalizations of the Heisenberg model, where deconfined critical points in the universality class of the $CP^{N-1}$ model have been found for $N>4$ \cite{KaulSandvik2012,PhysRevB.88.220408,kaul2015numerical}.

So far, most deconfined critical points in magnets that have been discussed in the literature are in the universality class of the $CP^{N-1}$ model. In this work we are going to study a scenario for a deconfined critical point in a different universality class. In particular we consider $SU(3)$ antiferromagnets in two dimensions, where a spin in the fundamental representation of $SU(3)$ is placed on each site of a triangular lattice (note that this is in contrast to the $SU(N)$ generalizations in Ref.~\cite{KaulSandvik2012}, where spins on the two different sublattices of the bipartite square lattice transform under fundamental and conjugate representations, respectively).
Such $SU(3)$ antiferromagnets appear at a specific parameter point of the more general spin-1 bilinear-biquadratic (BBQ) model \cite{papanicolaou1988unusual}. Moreover, they can be realized in systems of ultracold atoms, where they arise in the strong coupling Mott limit of $SU(3)$ symmetric Hubbard models with $3$ flavors of fermions. The physics of such $SU(N)$ magnets, which host a multitude of novel states, have been realized in several 
cold atom setups in recent years \cite{honerkamp2004ultracold,
gorshkov2010two, taie2010realization, 
taie20126, hofrichter2016direct}. 
In our work we study possible continuous transitions between $SU(3)$ analogues of the magnetically ordered N\'eel phase and the VBS phase. We argue that the critical theory can be written in terms of three coupled $CP^2$ models, which features a new critical fixed point.

The rest of this article is outlined as follows: In Sec.~\ref{allphasessec} we introduce the model and discuss the two symmetry broken phases of interest. A critical theory in terms of a $CP$-parametrization is constructed in Sec.~\ref{allcritical} starting from both, a non-linear sigma model description, and by fractionalizing an appropriate order parameter. We briefly discuss the mean field phase diagram as well. In Sec.~\ref{FRG-Analysis} we perform a one-loop renormalization group study using the framework of functional renormalization group (FRG), where we treat the gauge sector within the background field formalism. Conclusions are presented in Sec.~\ref{conclusion-section} and in appendix \ref{FRGapp} we outline details of the FRG computation. Finally, in appendix \ref{VBStheorysec} we present a complementary route to obtain the critical field theory, by analyzing topological excitations in the trimerized $SU(3)$ singlet phase.

\section{$SU(3)$ Antiferromagnets}
\label{allphasessec}

\begin{figure}
\centering
\includegraphics[width=0.7\columnwidth]{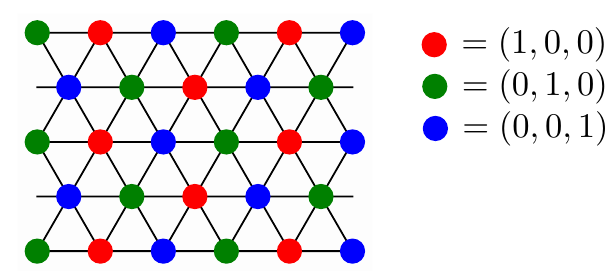}
\caption{A realization of the three-subattice color- ordered state of $SU(3)$ spins on the triangular lattice.}
\label{colororder}
\end{figure}

We consider an antiferromagnetic Heisenberg model on the triangular lattice with a spin in the fundamental representation of $SU(3)$ on each lattice site. Its Hamiltonian is given by 
\begin{align} 
\label{Heisenbergfirst}
H = J \sum_{\braket{i,j}} \boldsymbol{\lambda}_i\boldsymbol{\lambda}_j \ , \quad J>0 \quad ,
\end{align}
where $\boldsymbol{\lambda}_i$ is the eight-dimensional vector of Gell-Mann matrices, which are the generators of $SU(3)$, and the sum extends over nearest neighbors. The Hilbert space at lattice site $i$ is the projective space $CP^2$ of three-dimensional complex normalized vectors $\textbf{z}_i$ defined up to a phase (for brevity, we will call them ``spinors" in the following).
Defining
$\mathbf{m}_i = \braket{\boldsymbol{\lambda}_i}$, the mean field ground state is the well known $120^\circ$ ordered state where
\begin{align}
\label{120grad}
\sum_{i \in \bigtriangleup} \mathbf{m}_i =0  , \ \quad \norm{\textbf{m}_i} = \frac{2}{\sqrt{3}} \ ,
\end{align}
for every elementary plaquette $\bigtriangleup$ of the triangular lattice. Its mean field energy is found to be $-2J$ \cite{lauchli2006quadrupolar}. The $SU(3)$ flavor vectors $\mathbf{m}_i$ on the three sublattices are coplanar and span $120^\circ$ angles, while the corresponding complex spinors $\textbf{z}_i$ on the three sublattices are mutually orthogonal. 
Note that this configuration is the direct analogue of the $SU(2)$ N\'eel state on the square lattice. Indicating the basis vectors of $CP^2$ with colors red, green, blue, one possible realization of this color-ordered state is pictorially shown in Fig.\ \ref{colororder}.
In analogy to the staggered magnetization for ordinary square lattice $SU(2)$ antiferromagnets, we can define a scalar order parameter for the color-ordered phase by
\begin{align}
\label{nicen}
m_c = \norm{ \sum_{i\in A} \mathbf{m}_i + e^{i2\pi/3}\sum_{i\in B} \mathbf{m}_i + e^{i4\pi/3}\sum_{i\in C} \mathbf{m}_i } \ ,
\end{align}
where $A,B,C$ are the three sublattices. One can straightforwardly show that $m_c$ is maximized in the color-ordered state out of all possible states.

\begin{figure}
\centering
\includegraphics[width=0.7\columnwidth]{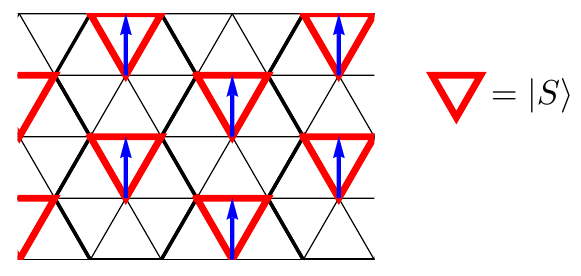}
\caption{A particular realization of the trimerized $SU(3)$ singlet state. The arrows represent the $\mathbb{Z}_6$-clock order parameter.}
\label{VBS}
\end{figure}

Numerical studies of the Hamiltonian in Eq.~\eqref{Heisenbergfirst} show that the exact ground state indeed exhibits three-sublattice color-order \cite{bauer2012three}. Generalized Hamiltonians with additional ring-exchange terms, which arise naturally from higher order terms in the usual strong-coupling expansion of the $SU(N)$ Hubbard model, have been studied in Refs.~\cite{bieri2012paired,lai2013possible}, where a variety of non-trivial paramagnetic ground states were found. Among several quantum spin liquid states, a trimerized $SU(3)$ singlet state was found in Ref.~\cite{lai2013possible}. This state is an analogue of the VBS state in $SU(2)$ spin systems. In the trimerized state the lattice is covered with $SU(3)$ singlets $|S\rangle$ formed by three spins on an elementary triangle 
\begin{align}
\ket{S} = \epsilon_{\alpha\beta\gamma} \, z_{\alpha 1} z_{\beta 2} z_{\gamma 3} \ ,
\end{align}
where $\varepsilon_{\alpha \beta \gamma}$ is the fully antisymmetric tensor of $SU(3)$ and a summation convention is used for greek indices. 
The singlets order in a specific pattern, thereby breaking lattice translation and rotation symmetries. 
Here, we will focus on the most simple singlet configurations, which can be indexed by a $\mathbb{Z}_6$-clock order parameter. They correspond to a six-fold degenerate paramagnetic ground state.
A pictorial representation of a trimer state is shown in Fig.~\ref{VBS}.

In this work we want to address the question whether a direct second order quantum phase transition between the color-ordered and the trimerized phase is possible, and study its properties. This transition would be a direct analogue of the deconfined critical point for the N\'eel-VBS transition in $SU(2)$ antiferromagnets on the square lattice.

\section{Critical Theory}
\label{allcritical}

\subsection{Path-integral derivation of the critical theory}
\label{Pathintsec}

Our critical theory will be based on the non-linear sigma model for the color ordered state derived by Smerald and Shannon \cite{smerald2013theory}, which we briefly review in the following. 
The starting point is the $SU(3)$-symmetric point of the bilinear-biquadratic model, given by 
\begin{align}
\label{BBQ}
H_{\text{BBQ}}^{SU(3)} =  J\sum_{\braket{i,j}} \mathbf{S}_i\cdot \mathbf{S}_j + \left(\mathbf{S}_i\cdot \mathbf{S}_j\right)^2 \ ,
\end{align}
where $\mathbf{S}_i$ are spin-1 operators. 
Up to a constant shift and rescaling of $J$, this Hamiltonian is equivalent to Eq.~\eqref{Heisenbergfirst} \cite{aguado2009density}.
In terms of spin-1 coherent states Eq.~\eqref{BBQ} can then be rewritten as
\begin{align}
H_{\text{BBQ}}^{SU(3)} =  J \sum_{\braket{i,j}} |\textbf{d}_i \cdot \bar{\textbf{d}}_j|^2 \ ,
\end{align}
where $\textbf{d}_i$ is a three dimensional complex normalized vector. In fact, the overall phase of $\textbf{d}_i$ is fixed in Ref.~\cite{smerald2013theory}, but the precise phase choice is immaterial at this stage of the analysis.
In the next step, the fluctuating fields $\textbf{d}_i$ are expanded around a generic color-ordered configuration, which is parametrized by three mutually orthogonal, complex vector fields $\textbf{z}_1 ,\textbf{z}_2,\textbf{z}_3$, where $1,2,3$ correspond to the three sublattices. These fields fulfill 
\begin{align}
\label{orthogonality constraint}
\bar{\textbf{z}}_i \cdot \textbf{z}_j = \delta_{ij} \ ,
\end{align}
Deviations from this color-order are parametrized in terms of small "canting" fields $\textbf{l}$, which can be integrated out at the quadratic level. Changing to a continuum description and introducing a kinetic term (which arises in the standard way from the path-integral construction), the resulting partition function in the zero temperature limit is given by 
\begin{align}
\label{somepartitionfct}
Z\sim \int \displaystyle{\prod_i} \tilde{\mathcal{D}} \z_i \  \prod_{j\geq i} \delta(\bar{\textbf{z}}_i \cdot \textbf{z}_j - \delta_{ij}) \exp(- S) \quad \ ,  
\end{align}
where the measure $\tilde{\mathcal{D}}\z_i$ contains a gauge fixing of the phase of $\z_i$ to avoid double counting of physical degrees of freedom. The euclidean action $S$ appearing in (\ref{somepartitionfct}) reads 
\begin{align}
\label{action10}
S = \int d^3x\  \alpha \sum_i \bar{\z}_i\cdot \partial_\tau \z_i + \sum_{\mu, i\neq j} |\bar{\z}_i\cdot \partial_\mu \z_j|^2 \quad ,   
\end{align}
where $\mu \in \{0,1,2\}$, and $\tau = x_0$ denotes the imaginary time direction. Here $\alpha$ is a numerical coefficient which depends on $J$ and the details of the continuum limit, and will not be of importance to us.

An important property of the action in Eq.~\eqref{action10} is its manifest invariance under \textit{sublattice-dependent} $U(1)$ gauge transformations of the form
\begin{align}
\label{gaugetrafo}
\z_i(\x) \rightarrow e^{i\theta_i(\x)}\z_i(\x) \ .
\end{align}
For the second term in Eq.~(\ref{action10}) this invariance follows from the orhogonality constraint contained in Eq.~(\ref{orthogonality constraint}), while the first term only picks up a total time-derivative under gauge transformation by virtue of the normalization constraint.

The first term in Eq.~(\ref{action10}) corresponds to a topological Berry phase term. 
Only singular field configurations should give a nonzero Berry-phase contribution. 
By analogy to $SU(2)$, we can expect these configurations to be hedgehog-events, where an appropriately defined soliton winding number jumps in time. In the $SU(2)$ case on the square lattice, inclusion of Berry phase terms renders these events dangerously irrelevant at the quantum critical point, but relevant in the paramagnetic phase \cite{senthil2004quantum}. In the gauge language, the hedgehogs correspond to magnetic monopoles, and their irrelevance makes the resulting $U(1)$ gauge theory non-compact. 
For $SU(3)$, the soliton-structure on the color-ordered side was recently studied in Ref.~\cite{ueda2016quantum} by a homotopy analysis of the ground state manifold, giving rise to a $\mathbb{Z}\times \mathbb{Z}$ winding number classification. We will not perform an analysis of the corresponding hedgehog events here and disregard the Berry-phase terms altogether, assuming that their only role is to render the $U(1)$ gauge field compact as in the $SU(2)$ case.  

Following Refs.~\cite{ueda2016quantum,Auerbach1994} we can bring the remaining action in another form by introducing the following real functions of the $\z_i$-fields 
\begin{align}
\tilde{\mathcal{A}}^i_\mu = -\frac{i}{2} \left[\bar{\z}_i\partial_\mu \z_i - (\partial_\mu \bar{\z}_i) \z_i\right]
\end{align}
Under the gauge transformation in Eq.~\eqref{gaugetrafo} $\tilde{\mathcal{A}}^i_\mu$ transforms as 
\begin{align}
\label{atrafo}
\tilde{\mathcal{A}}^i_\mu \rightarrow \tilde{\mathcal{A}}^i_\mu + \partial_\mu \theta_i \ .
\end{align}
With help of these fields, and the identities in Eq.~\eqref{orthogonality constraint}, the Lagrangian is now rewritten as (c.f. \cite{ueda2016quantum})
\begin{align} \sum_{\mu, i\neq j} |\bar{\z}_i\cdot \partial_\mu \z_j|^2 = \sum_{i,\mu} |\partial_\mu \z_i|^2 - \big(\tilde{\mathcal{A}}^i_\mu\big)^2\ .
\end{align}
Following Ref.~\cite{Auerbach1994} we may trade the $\z$-dependent gauge fields $\tilde{\mathcal{A}}$ for $\z$-independent gauge fields $\mathcal{A}$ with help of a Hubbard-Stratonovich transformation of the form 
\begin{align}
\label{HStrafo}
\exp\left[{\tilde{\mathcal{A}_\mu^i}^2}\right] \sim \int_{-\infty}^\infty d\mathcal{A}_\mu^i \exp\left(-\mathcal{A}_\mu^i + 2\mathcal{A}_\mu^i\tilde{\mathcal{A}}_\mu^i \right) \ .
\end{align}
For Eq.~\eqref{HStrafo} to hold after gauge transformations, the fields $\mathcal{A}$ must inherit the transformation properties of the fields $\tilde{\mathcal{A}}$ given by (\ref{atrafo}). Inserting (\ref{HStrafo}) into the partition function, we finally arrive at
\begin{align}\nonumber
Z = \int &\displaystyle{\prod_{i,\mu}} \mathcal{D} \z_i  \mathcal{\tilde{D}}\mathcal{A}^i_\mu \  \prod_{j\geq i} \delta(\bar{\textbf{z}}_i \cdot \textbf{z}_j - \delta_{ij}) \\  &\exp\left\{-\int d^3 x \sum_{i,\mu} | \left(\partial_\mu - i\mathcal{A}^i_\mu\right)\z_i|^2\right\} \ .
\label{intermediateZ}
\end{align}
In writing Eq.~(\ref{intermediateZ}) we tacitly performed the following manipulation: the gauge fixing term for the phase of the $\z$-fields, which was contained in the measure $\mathcal{\tilde{D}}\z$ in Eq.~(\ref{somepartitionfct}), is carried over to an equivalent gauge fixing condition in the measure $\tilde{\mathcal{D}}\mathcal{A}$ in Eq.~(\ref{intermediateZ}), which will be made explicit by introducing a standard gauge fixing term later.

The action of Eq.~(\ref{intermediateZ}) bears close resemblance to the $CP^1$-action of Eq.~(\ref{cp1model}). In fact, one may imagine a derivation of Eq.~(\ref{cp1model}) for $SU(2)$ largely analogous to the one presented above, with two-dimensional $\textbf{z}_i$-fields, and two sublattices only. However, for $\z_i \in \mathbb{C}^2$, the orthogonality constraint contained in Eq.~(\ref{intermediateZ}) fully determines $\z_2$ as function of $\z_1$ (or vice versa), up to a phase. This can be made explicit by writing 
\begin{align} 
\label{SU2propconstraint}
z_{\alpha 2} \sim \epsilon_{\alpha \beta} \bar{z}_{\beta 1} \ .
\end{align}
Inserting this into (\ref{intermediateZ}), the gauge fields couple to the same $\z$ in identical fashion and are indistiguishable. Softening the unit length constraint on $\z$, one therefore recovers (\ref{cp1model}) (up to the Maxwell term, which is generated during the RG flow, see below).

\subsection{Fractionalizing the order parameter}

For spin-1/2 models, a common shortcut in deriving critical actions is a fractional parametrization of an appropriate real order parameter \cite{sachdev2011quantum,chubukov1994quantum}, as shown in Eq.~\eqref{zmappingsu2} for the N\'eel state.
We can proceed accordingly, identifying the triple of physical 8-D flavor vectors $\textbf{m}_i$ with  $120^\circ$ order from Eq.~\eqref{120grad} as order parameter. The vectors $\textbf{m}_i$ parametrize the manifold of classical ground states, which are, in fact, product states. Therefore, we can re-express them as 
\begin{align}
\label{ourspinor}
\textbf{m}_i = \bar{z}_{\alpha i} \boldsymbol{\lambda}_{\alpha\beta} z_{\beta i} \ , \qquad i,\alpha,\beta \in \{1,2,3\} \  , 
\end{align}  
where the fields $\z_i$ precisely fulfill the normalization and orthogonality constraints of Eq.~\eqref{orthogonality constraint}. The phase ambiguity in Eq.~\eqref{ourspinor} entails a three-fold gauge invariance, and an action in terms of $\z_i$ can be readily derived by expanding in covariant derivatives. This reasoning immediately gives us our critical theory in Eq.~\eqref{intermediateZ}. One should note that, while the above argumentation seems to be limited to the description of classical order parameter fluctuations, the previous path integral formulation explicitly shows that the critical theory does include quantum fluctuations as well.

\subsection{Softening the constraints}
\label{constraintsoftsec}
The critical theory in Eq.~\eqref{intermediateZ} is rather inconvenient to handle due to the delta-function constraints. We can proceed by softening the constraints and replace them by an appropriate potential $V(\z)$, which must obey the following properties: 
(i) invariance under global $SU(3)$-rotations $\z_i\rightarrow \hat{U}\z_i$;
(ii) invariance under lattice symmetries, which simply permute the sublattice indices (e.g.~under rotations with base point on sublattice $1$: $\z_1 \rightarrow \z_1, \z_2 \leftrightarrow \z_3$);
(iii) $U(1)$ gauge invariance. 
Expanding up to quartic terms, the resulting general potential has the form 
\begin{align}
\label{generalV}& V(\z) =  \sum_{i}\left\{m \left(\z_i\cdot \bar{\z}_i\right)  +  \rho_1 (\z_i\cdot \bar{\z}_i)^2 \right\} +  \\ \notag &  2\sum_{i \neq j}\left\{ \rho_2 (\z_i\cdot \bar{\z}_i)(\z_j \cdot\bar{\z}_j)  + \rho_3(\z_i\cdot \bar{\z}_j)(\z_j \cdot\bar{\z}_i) \right\} \ , 
\end{align}
where the factor of $2$ is introduced for later convenience, and $m, \rho_1, \rho_2, \rho_3$ are real coupling constants. Our resulting critical theory therefore reads 
\begin{align}
\label{finalaction}
S = \int d^3 x \sum_{i,\mu} | \left(\partial_\mu - i\mathcal{A}^i_\mu\right)\z_i|^2 + V(\z)  \ ,
\end{align}
which features three $CP^2$-models coupled via quartic interaction terms.

\begin{figure}
\centering
\includegraphics[width=\columnwidth]{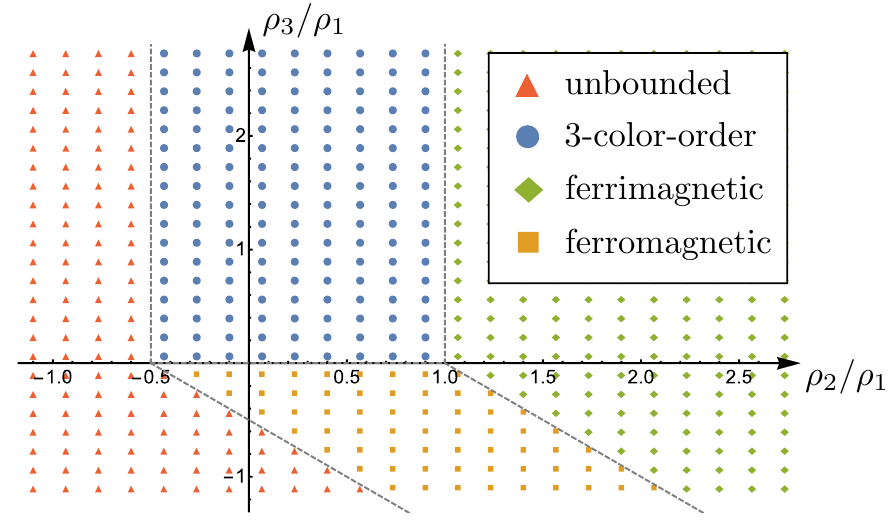}
\caption{Mean field phase diagram corresponding to the potential $V(\z)$ from Eq.~\eqref{generalV}. The dashed lines indicate the phase boundaries.}
\label{meanfieldphasediag}
\end{figure}

To gain some insight into the structure of $V$, let's perform a mean field analysis, restricting ourselves to the ordered phase where $m<0, \rho_1>0$. It is obvious that the term $\sim  \rho_3$ is the only one which depends on the relative direction of the spinors: When $\rho_3 > 0$, the spinors are orthogonal on the mean field level, and parallel for $\rho_3<0$.
Some easy algebra then yields the following mean field boundaries: 
First, for $\rho_3>0$: 
\begin{itemize} 
[labelindent=0pt,labelwidth=\widthof{\ref{last-item1}}, itemindent=1.4em,leftmargin=!]
\item[--] $\rho_2 < - \frac{1}{2} \rho_1$: For this (unphysical) parameter choice, the potential is not bounded below. 
\item[--] $- \frac{1}{2} \rho_1<\rho_2 < \rho_1 $: These values correspond to a well-defined three color order, with nonzero expectation values and mutual orthogonality for all $\z_i$. 
\item[--] $\rho_2 > \rho_3$: One finds a "ferrimagnetic" phase, where the expectation value of two spinors $\z_i$ is zero.  
\label{last-item1}
\end{itemize}
Second, for $\rho_3<0$: 
\begin{itemize} 
[labelindent=0pt,labelwidth=\widthof{\ref{last-item2}}, itemindent=1.4em,leftmargin=!]
\item[--] $\rho_3 < - \frac{1}{2} \rho_1 - \rho_2$: The potential is unbounded.
\item[--] $- \frac{1}{2} \rho_1-\rho_2<\rho_3 <\rho_1-\rho_2$: Corresponds to a "ferromagnetic" phase, where all spinors have a nonzero expectation value and point in the same direction. 
\item[--] $\rho_1 -\rho_2 > \rho_3$: Ferrimagnetic phase.   \label{last-item2}
\end{itemize}
The above phases are summarized in Fig.\ \ref{meanfieldphasediag}.

\section{FRG-Analysis}
\label{FRG-Analysis}

\subsection{General FRG setup}

We can analyze the possible phase transitions predicted by the critical action $S$ by looking for its RG fixed points. To derive the flow equations, our method of choice is Functional Renormalization Group (FRG), employing conventions from \cite{Kopietz2010,bartosch2013corrections}.
The backbone of this analysis is the functional  
\textit{Wetterich equation}, which describes the running of the scale-dependent Legendre effective action $\Gamma_\Lambda$ under variation of the momentum cutoff-scale $\Lambda$. At the initial UV scale $\Lambda_0$, $\Gamma_{\Lambda_0}$ reduces to the bare action $S$; for $\Lambda \rightarrow 0$, $\Gamma_\Lambda$ becomes the Legendre transform of the true generating functional of connected Green's functions. This is achieved by successively integrating out UV degrees of freedom via inclusion of regulator terms $\mathcal{R}_\Lambda$, which suppress IR fluctuations.
 Taking functional derivatives, the flow of $\Gamma_\Lambda$ can then be projected on the flow of the coupling constants. 
  
While the FRG treatment of the scalar sector is very straightforward, technical difficulties arise upon including gauge degrees of freedom. Several workarounds are available \cite{gies2012introduction}; following previous treatments of $CP^n$-models \cite{reuter1993average, reuter1994exact,reuter1993running,bergerhoff1996phase, bartosch2013corrections}, we will employ the \textit{background field} formalism, introduced by Reuter and Wetterich. 
Its main idea is to work with an effective action $\Gamma_\Lambda$ which is manifestly gauge invariant, while at the same time containing a gauge-fixing term necessary for well-defined functional integrals in the first place. This gauge-invariant formulation allows to choose a meaningful truncation of $\Gamma_\Lambda$, as necessary to make any technical progress.

To implement this idea, one first expands the dynamical gauge fields $\mathcal{\A}$ appearing in the bare action around some fixed background field $\bar{\A}$ (we suppress indices for now), which gauge-transforms in the standard way. The effective action $\Gamma_\Lambda$, obtained via an appropriate Legendre transform of $S$, then depends on $\A = \braket{\boldsymbol{\mathcal{A}}}, \phi = \braket{z}$, and $\bar{\A}$ (averages are taken w.r.t.\ to $S$ along with sources and regulator terms), and is gauge invariant under gauge-transformations of both $\bar{\A}, \A$ and $\phi$.    
However, since arbitrary powers of $(\bar{\A} - \A)$ are gauge invariant, using $\Gamma_\Lambda[\A,\bar{\A}, \phi]$ is still inconvenient, and one needs to eliminate the field $\bar{\A}$. This can be achieved by identifying it with $\A$. In doing so, one picks up spurious functional derivatives, which can be partially accounted for by an appropriate gauge-invariant counterterm $\mathcal{C}_\Lambda(\A)$. This term will modify the flow equation of the gauge coupling only.
Ultimately, defining an appropriate effective action, one arrives at the following approximate flow equation 
\begin{align}
\label{Wetterich}
&\frac{\partial}{\partial l}\Gamma_\Lambda[\phi,\A] \simeq \\ \nonumber
& \frac{1}{2} \text{Tr}\left[\frac{\partial}{\partial l} \mathcal{R}_\Lambda[\A]\left(
\Gamma_\Lambda^{(2)}[\phi,\A] + \Gamma_\text{gf}^{(2)} + \mathcal{R}_\Lambda[\A] \right)^{-1}  \right] \\ \nonumber &
+ \frac{\partial}{\partial l}\mathcal{C}_\Lambda[\A] \ .
\end{align}
Here 
$l$ is the logarithmic RG scale, connected to the momentum cutoff-scale $\Lambda$ by 
\begin{align}
\Lambda = \Lambda_0 \exp(-l) \ .
\end{align} 
$\Gamma_\Lambda^{(2)}$ and $\Gamma^{(2)}_{\text{gf}}$ are second derivatives of $\Gamma_\Lambda$ and the gauge-fixing term $\Gamma_{\text{gf}}$ (we choose the Lorenz gauge) w.r.t.\ the fields $\A, \phi$. All objects on the r.h.s.\ of (\ref{Wetterich}) are matrix-valued in $(\phi,\bar{\phi}, \A)$-space, and the trace involves a summation in this space as well as over all internal indices.

\begin{figure}
\centering
\includegraphics[width = \columnwidth]{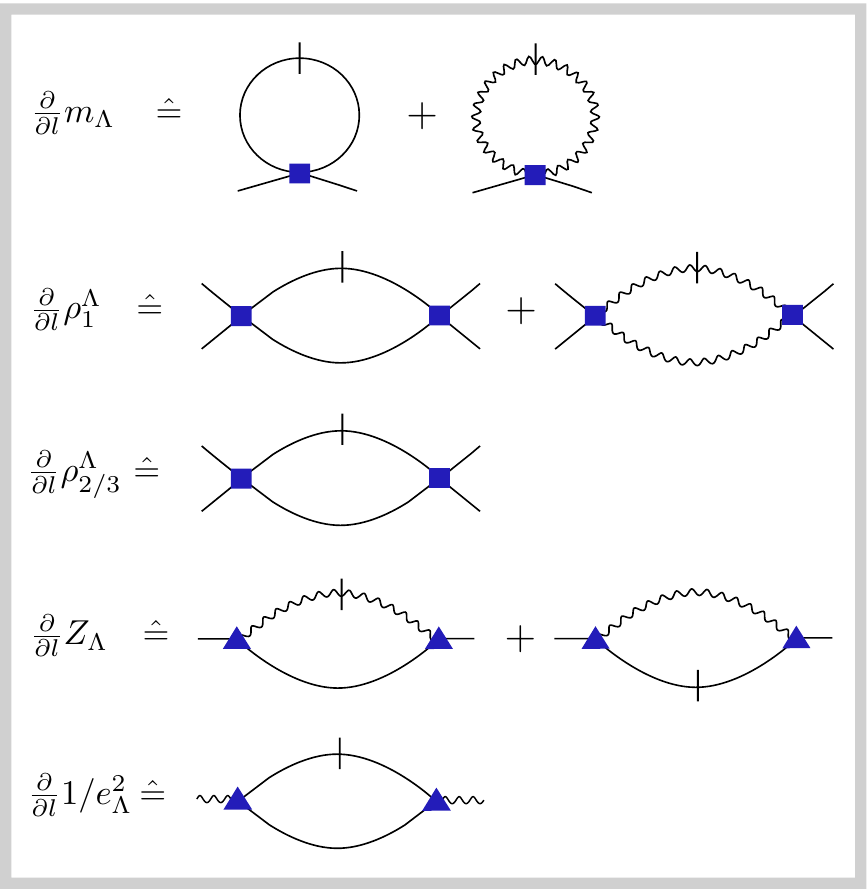}
\caption{Diagrammatic representation of the flow equations. Full (wiggly) lines denote scalar (gauge) field propagators, while blue square (triangles) denote 4-point (3-point) vertices. Vertical lines represent regulator insertions.}
\label{diagrampics}
\end{figure}

To proceed, we need to specify an ansatz for an effective action $\Gamma_\Lambda$. We choose it to be equal to the bare action, with running couplings, a standard wave-function renormalization term $Z_\Lambda$, and lattice-dependent Maxwell terms coupled by a running gauge charge $e_\Lambda$.  
Explicitly writing out the index structure and employing Einstein sums we have 
\begin{align} \nonumber
\label{effectiveAnsatz}
&\Gamma_\Lambda[\A,\phi]= \int\! d^3 x\bigg\{  Z_\Lambda| (\partial_\mu -i A^i_\mu)\phi_{\alpha i}|^2 + m_\Lambda \phi_{\alpha i}\bar{\phi}_{\alpha i} \\  & 
+ \hat{\rho}^\Lambda_{ijkl}{\phi}_{\alpha i}\bar{\phi}_{\alpha j}\phi_{\beta k}\bar{\phi}_{\beta l} + \frac{1}{4e^2_\Lambda} (\mathcal{F}^i_{\mu\nu})^{2} \bigg\}\ , 
\end{align}
where $\hat{\rho}^\Lambda$ is a compact notation for the quartic potential terms of the form
\begin{align}
\hat{\rho}^\Lambda_{ijkl} =  
\begin{cases}
\rho^\Lambda_{1}, \quad i=j=k=l\\
\rho^\Lambda_{2},\quad  i=j, k=l, i\neq k\\
\rho^\Lambda_{3}, \quad i\neq j, i= l, k=j \\
0 \qquad \text{else} \ . 
\end{cases}
\end{align}
Roman letters always denote sublattice indices, $\alpha,\beta,\gamma$ $SU(3)$ indices, and $\mu, \nu$  spacetime indices.  
One should note that all gauge charges are equal, as required by translational invariance, which permutes the gauge field sublattice indices. Also, we disregared terms which couple the gauge fields on different sublattices to each other, since they renormalize on two-loop level only.
Specifying the regulator $\mathcal{R}_\Lambda[\A]$ and the counterterm $\mathcal{C}_\Lambda[\A]$, we can obtain the one-loop flow equation of the couplings from the central equation (\ref{Wetterich}) by taking appropriate functional derivatives w.r.t.\ the fields. 
A pictorial representation of these flow equations is shown in Fig. \ref{diagrampics}.

Unlike earlier studies of $CP^n$-models, we take the functional derivatives at zero scalar fields for simplicity, approaching the fixed point from the symmetric phase. This usually leads to numerically less precise results for the critical exponents, but is sufficient to determine the fixed point structure of our theory. To correct for these truncation related errors, we derive the flow equations
for general $SU(N)$, i.e.~we extend the summations over the three $SU(3)$ indices $\alpha, \beta$ in Eq.~\eqref{effectiveAnsatz} to a summation which ranges from $1$ to $N$. 
We then study the behaviour of the flow equations in the large $N$ limit. In the $CP^{N-1}$-case, this was shown to yield qualitatively correct results in Ref.~\cite{bergerhoff1996phase}.

Technical details of the derivation are given in appendix \ref{FRGapp}. We phrase the flow equations in terms of dimensionless quantities 
\begin{align}
\label{rescaledquant}
\tilde{m}= \frac{m_\Lambda}{\Lambda^2 Z_\Lambda} \ , \quad 
\tilde{\rho}_i= \frac{\rho_i^\Lambda}{\Lambda  Z_\Lambda^2} \ , \quad 
\tilde{e}^2= \frac{e^2_\Lambda}{\Lambda} \ .  
\end{align}
Furthermore, we introduce the anomalous dimension of the scalar fields as 
\begin{align}
\label{anomdim}
\eta \equiv \frac{\partial}{\partial l} \log(Z_\Lambda).
\end{align}
The flow equations obtained this way read
\begin{widetext}
\begin{align}
\label{mfloweq}
\dl \tilde{m} &= \tilde{m}\cdot (2-\eta) + \frac{\tilde{e}^2 \left(\tilde{m}+1\right)^2+2 \left((N+1) \tilde{\rho }_1+2 \left(N
   \tilde{\rho }_2+\tilde{\rho }_3\right)\right)}{3 \pi ^2
   \left(\tilde{m}+1\right)^2} \\
\label{rho1floweq}
\dl \tilde{\rho}_1 &= \tilde{\rho}_1 \cdot (1-2\eta)
-\frac{4 \left(\tilde{e}^4 \left(\tilde{m}+1\right)^3+(N+4) \tilde{\rho }_1^2+2
   \left(N \tilde{\rho }_2^2+2 \tilde{\rho }_3 \tilde{\rho }_2+\tilde{\rho
   }_3^2\right)\right)}{3 \pi ^2 \left(\tilde{m}+1\right)^3}
\\
\label{rho2floweq}
\frac{\partial}{\partial l} \tilde{\rho}_2 &=\tilde{\rho}_2 \cdot (1-2\eta) 
-\frac{4 \left((N+2) \tilde{\rho }_2^2+2 \tilde{\rho }_1 \left((N+1) \tilde{\rho
   }_2+\tilde{\rho }_3\right)+2 \tilde{\rho }_3 \tilde{\rho }_2+\tilde{\rho
   }_3^2\right)}{3 \pi ^2 \left(\tilde{m}+1\right)^3}
\\
\label{rho3floweq}
\frac{\partial}{\partial l} \tilde{\rho}_3 &=\tilde{\rho}_3 \cdot (1-2\eta) 
-\frac{4 \tilde{\rho }_3 \left((N+1) \tilde{\rho }_3+2 \tilde{\rho }_1+4 \tilde{\rho
   }_2\right)}{3 \pi ^2 \left(\tilde{m}+1\right)^3}
\\
\label{eflowfinaleq}
\frac{\partial}{\partial l} \tilde{e}^2 &= \frac{\tilde{e}^2 \left(12 \pi ^2 \left(\tilde{m}+1\right)-N \tilde{e}^2\right)}{12 \pi ^2 \left(\tilde{m}+1\right)} + 
\frac{\tilde{e}^4 N \left(\tilde{m} \sqrt{\tilde{m}+2}-2 (\tilde{m}+1) \coth ^{-1}\left(\sqrt{\tilde{m}+2}\right)\right)}{12 \pi ^2 (\tilde{m}+1) (\tilde{m}+2)^{3/2}} \\ 
\label{etafloweq}
\eta &= -\frac{8 \tilde{e}^2 \left(\tilde{m}+2\right)}{9 \pi ^2 \left(\tilde{m}+1\right)^2} \ .
\end{align}
\end{widetext}

\subsection{Fixed point structure}

Let's analyze the fixed point structure of the above flow equations. The fixed points are obtained by numerically solving for the zeros of the beta functions. Linearizing the beta functions around the fixed points and determining the eigenvalues of the resulting coefficient matrix, one can then derive the stability properties. Note that the coefficient matrix is in generally not symmetric (s.t.\ the left and right eigenvectors do not coincide), but always found to be diagonalizable.

As sanity check, one can set $\tilde{\rho}_2 = \tilde{\rho}_3 = 0$ (which is of course a solution to Eqs.~\eqref{rho2floweq}, \eqref{rho3floweq}). Then, our model simply reduces to 3 copies of the standard $CP^{N-1}$ model, and we can compare the fixed point structure to prior treatments, in particular to the large $N$ analysis of Bergerhoff et al.\  \cite{bergerhoff1996phase}. 
For small or moderate $N$, as can be deduced from Eq.~\eqref{eflowfinaleq}, no fixed point at nonzero gauge charge is found, and the flow equations reduce to the ones of the usual $O(2N)$-model: The gauge field fluctuations are not strong enough to renormalize the scalar sector. 
As a result, there are just two fixed points: the Gaussian and the Wilson-Fisher fixed point, which has one additional unstable direction corresponding to the gauge charge. Since the scalar field anomalous dimension is exclusively generated by gauge field fluctuations within our treatment ($\eta \sim \tilde{e}^2$), in this regime we find $\eta = 0$. 

The picture changes for $N \geq 353$, where two further fixed points at non-zero gauge charge appear, corresponding to additional zeros of the gauge charge beta function (\ref{eflowfinaleq}).
The analytical structure of these addtional fixed points is transparent in the limit $N\rightarrow \infty$: First, one observes that the gauge flow equation (\ref{eflowfinaleq}) has the form
\begin{align}
\label{esquared form}
\frac{\partial}{\partial l} \tilde{e}^2 = \tilde{e}^2 \cdot \left(1-N f(\tilde{m})\cdot \tilde{e}^2\right) \ , 
\end{align}
where $f(\tilde{m})$ is some function; for $\tilde{m}$ sufficiently larger than $-1$, which is always fullfilled for meaningful fixed points, $f(\tilde{m})$ is of order $1$. Therefore, the fixed point value $\tilde{e}^2_\star$ scales as $1/N$. Linearizing (\ref{esquared form}), the corresponding RG-eigenvalue is $-1$.

Since $\tilde{e}^2_\star \sim 1/N$, the leading $N$ behaviour of the scalar sector near the fixed points then completely decouples from the gauge sector, and the scalar flow equations reduce to: 
\begin{align}
\label{mflowlargeN}
\dl \tilde{m} &= 2\tilde{m} + \frac{2}{3\pi^2} N \cdot  \frac{\tilde{\rho}_1}{\left(1+\tilde{m}\right)^2} + \mathcal{O}\left(\frac{1}{N}\right)\\
\label{rhoflowlargeN}
\dl \tilde{\rho}_1 &= \tilde{\rho}_1 -\frac{4}{3\pi^2}N\cdot  \frac{ \tilde{\rho}_1^2}{\left(1+\tilde{m}\right)^3} + \mathcal{O}\left(\frac{1}{N^2}\right) \ .
\end{align}
These equations have the usual Gaussian and Wilson-Fisher fixed points. At $\tilde{e}^2_\star >0$, the Gaussian fixed point of the scalar sector gives rise to the ``tricritical fixed point'' \cite{Bergerhoff1996} of the full theory, while the Wilson-Fisher fixed point of the scalar sector determines the critical fixed point of the full theory; the leading large $N$ values of the latter read $\tilde{m}_\star = -1/5, \tilde{\rho}_{1\star} = 48\pi^2/(125N)$, with corresponding eigenvalues $\sqrt{5/2}, -\sqrt{5/2}$ \footnote{Note that the textbook eigenvalues (-1,1) \cite{sachdev2011quantum} found via $\epsilon$-expansion are recovered if we expand all equations in $\tilde{m}$}.
Focusing on the critical fixed point, we can also recover the subleading terms in $1/N$ from a numerical evaluation of the full flow equations. The relevant fixed point values are shown in tab.\ \ref{FPtable} (right column), and the RG eigenvalues obtained are presented in tab.\ \ref{RGEVtable} (right column). Note that the RG eigenvalues correspond to scaling operators which are linear combinations of the original couplings. 

While the qualitative $N$-dependence of all relevant quantities coincides with the findings of \cite{Bergerhoff1996}, their numerical values are rather distinct. E.g.\ for the anomalous dimension $\eta$, our result $\simeq -14/N$ is quite different from the result $-0.31/N$ given by \cite{Bergerhoff1996}. This can be attributed to the fact that we disregarded any contribution to $\eta$ arising from the scalar sector, since we approach the fixed point from the symmetric phase. As we merely want to clarify if a stable fixed point exists for our full theory, we will overlook these numerial deviations.
\begin{table}
\begin{ruledtabular}
\begin{center}
    \begin{tabular}{ccc}
       \multicolumn{1}{r}{}   & \multicolumn{1}{c}{\textbf{Full theory}} & 
   \multicolumn{1}{c}{$\boldsymbol{CP^{N-1}}$ \textbf{theory}} \\ 
 	  \hline
 	\multicolumn{1}{ c }{$\tilde{m}_\star$}
     &$-1/5+14/N$ & $-1/5+13/N$ \\ 
     \multicolumn{1}{ c }{$\tilde{\rho}_{1\star}$}
     & $48\pi^2/125\cdot 1/N -110/N^2$ & $48/125\pi^2 \cdot 1/N -120/N^2$ \\ 
     \multicolumn{1}{ c }{$\tilde{\rho}_{2\star}$}
    & $-144 \pi^2/125 \cdot 1/{N^2} $  & 0  \\
    \multicolumn{1}{ c }{$\tilde{\rho}_{3\star}$}
    & $48 \pi^2/125 \cdot 1/{N} + 290/N^2$  & 0  \\
    \multicolumn{1}{ c }{$\tilde{e}^2_{\star}$}
    & $54/{N}$ & $54/{N}$  \\
    \multicolumn{1}{ c }{$\eta$}
    & $-14/N $ & $-14/N$  \\
    \end{tabular}
    \caption{Fixpoint values of the stable fixed point (one relevant operator) for the full theory of the flow equations (\ref{mfloweq})--(\ref{eflowfinaleq})   (left column), and the $CP^{N-1}$ theory obtained by setting $\tilde{\rho}_2 = \tilde{\rho}_3 = 0$ (right column), to next to leading order in $1/N$. The last two quantities, being strongly truncation dependent, are given to leading order only. }
    \label{FPtable}    
\end{center}
\end{ruledtabular}
\end{table}
\begin{table}
\begin{ruledtabular}
\begin{center}
\begin{tabular}{cc}
 \multicolumn{1}{c}{\textbf{Full theory}} & 
   \multicolumn{1}{c}{$\boldsymbol{CP^{N-1}}$ \textbf{theory}} \\ 
 	  \hline
 	  $\sqrt{5/2}+20/N$ &  $\sqrt{5/2}+18/N$
 	\\ 
 	 $-\sqrt{5/2}+190/N$ &  $-\sqrt{5/2}+180/N$
 	\\ 
 	 $-1 + 90/N$ &  $-1+ 80/N$
 	\\ 
 	 $-1 - 24/N$ &  \ --
 	\\ 
     $-1 +90/N$ &  \ --  \\  	 
    \end{tabular}
\end{center}
\caption{RG eigenvalues of the critical fixed point, in next to leading order in $1/N$.}
\label{RGEVtable}
\end{ruledtabular}
\end{table}

Having discussed the $CP^{N-1}$-case we now return to the full theory in question, where $\tilde{\rho}_2, \tilde{\rho}_3 \neq 0$ in general. Solving for zeros of the flow equations numerically, we find a quite similar fixed point structure as before: For small $N$, four unstable fixed points with vanishing gauge charge are found. When $N$ is increased (above $N\gtrsim 20$), additional unstable zero gauge fixed points are found. For $N>353$, unstable fixed points at nonzero gauge charge appear. Finally, for $N \geq 382$, a stable critical fixed point is found, out of a total of $\mathcal{O}(15)$ fixed points. We can determine its large $N$ properties (semi-)analytically as follows: first, we perform a numerical large $N$ scaling analysis, which reveals 
the same scaling behaviour as in the $CP^{N-1}$-case for $\tilde{m}_\star, \tilde{\rho}_{1\star}, \tilde{e}^2_\star$, and furthermore $\tilde{\rho}_{2\star} \sim 1/N^2, \tilde{\rho}_{3\star} \sim 1/N$. We then insert this behaviour back into the flow equations, and keep the terms that are leading in $1/N$ only. This yields the following result: As before, the beta functions for $\tilde{m}, \tilde{\rho}_1$ decouple and take the forms (\ref{mflowlargeN}), (\ref{rhoflowlargeN}). The relevant solution is the Wilson-Fisher fixed point, with the same leading behaviour of the fixed point values and RG eigenvalues as before. The remaining large $N$ form of the flow of $\tilde{\rho}_2, \tilde{\rho}_3$ reads, upon inserting the critical values $\tilde{m}_\star, \tilde{\rho}_{1\star}$: 
\begin{align}
\dl \tilde{\rho}_2 &= - \tilde{\rho}_2 - \frac{2\tilde{\rho}_3}{N} - \frac{125 \tilde{\rho}_3^2}{48 \pi^2} + \mathcal{O}\left(\frac{1}{N^3}\right) \\
\dl \tilde{\rho}_3 &= \tilde{\rho}_3 - \frac{125 N \tilde{\rho}_3^2}{48 \pi^2}  + \mathcal{O}\left(\frac{1}{N^2}\right) \ ,
\end{align}
with relevant fixed point solution
\begin{align}
\tilde{\rho}^\Lambda_{2\star} = -\frac{144 \pi^2}{125 N^2} \ , \quad \tilde{\rho}_{3\star} = \frac{48\pi^2}{125 N} \ , 
\end{align}
and eigenvalues $(-1,-1)$.
The subleading behaviour is then again determined numerically, and is shown in Tabs.~\ref{FPtable}, \ref{RGEVtable} (left columns). As a result, we find that the scaling properties of the previous critical point of the $CP^{N-1}$-model are only slightly modified by the presence of the two additional irrelevant couplings $\tilde{\rho}_{2\star}$, $\tilde{\rho}_{3\star}$. Finally we can estimate the correlation length exponent $\nu$, which corresponds to the inverse of the relevant RG eigenvalue and takes the value $\nu = \sqrt{2/5} - \mathcal{O}(1/N) \simeq 0.63$ in the large $N$ limit.

In our simple truncation a stable fixed point only appears for sufficiently large $N$. This is a well known problem in the RG treatment of gauge theories, which are often plagued by runaway RG flows \cite{halperin1974first}. However, since the stable $CP^{N-1}$ fixed point does survive in the limit $N\rightarrow 2$ when a more involved truncation is used \cite{bergerhoff1996phase, bartosch2013corrections}, we conjecture that the same holds true for our theory in the interesting limit $N\rightarrow 3$.

\subsection{Interpretation of the critical fixed point}

As stated in the introduction, there is solid evidence that the critical point of the $CP^1$ model describes the phase transition between the N\'eel phase and the valence bond solid in square lattice systems. By analogy, it seems natural to associate the critical fixed point found above with a continuous phase transition between the 3-color-ordered state and a paramagnet, possibly the trimer state discussed in Sec.~\ref{allphasessec}. However, this immediate interpretation is hindered by the fact that our phase diagram in the magnetically ordered phase (see  Fig.\ \ref{meanfieldphasediag}) allows for $3$ different magnetically ordered phases. 

As an attempt to resolve this conundrum, we give the following argument: at least in the large $N$ limit, the values of the criticial fixed point fulfill the relations (see Tab.\ \ref{FPtable}) 
\begin{align}
\tilde{m}_\star<0, \quad \tilde{\rho}_{1\star} = \tilde{\rho}_{3\star} \gg |\tilde{\rho}_{2\star}| \ .
\end{align}
These parametric relations will also carry over to the non-rescaled coupling values (see Eq.\ (\ref{rescaledquant})), at every finite value of the  cutoff scale $\Lambda$. Comparing with the phase boundaries given in sec.~\ref{constraintsoftsec}, we therefore see that, at least on the mean field level, the bulk phase ``adjacent'' to the critical fixed point is indeed the 3-color-ordered phase.

At this point our analysis doesn't make any statement about the structure of the paramagnetic state. In particular, it is not obvious why the paramagnet should be of the trimerized $SU(3)$ singlet type. While for $SU(2)$ the ground state cannot be a trivial disordered paramagnet due to the Lieb-Schultz-Mattis-Hastings-Oshikawa theorem \cite{lieb1961two,hastings2004lieb,oshikawa2000commensurability}, to our knowledge no direct generalization of this theorem to $SU(3)$ magnets in two dimensions is available. In addition, the detailed analysis of Berry phase effects in the $SU(2)$ case shows that the paramagnetic phase breaks lattice symmetries as expected in the VBS phase, which is a strong argument in favor of the dQCP scenario. Even though we do not present an analysis of Berry phase effects for the $SU(3)$ problem in this work, we give a complementary derivation of our critical theory starting from the paramagnetic, trimerized VBS phase in App.~\ref{VBStheorysec}. The fact that the same critical theory describes the transition out of both ordered phases provides a strong argument that our theory indeed provides the correct description of the deconfined quantum critical point.

\section{Conclusions and Outlook}
\label{conclusion-section}

This paper explored the possibility of a deconfined quantum critical point in $SU(3)$-magnets on the triangular lattice. Guided by the analogy to $SU(2)$-magnets on the square lattice, we constructed a critical theory for the continuous transition between a magnetically ordered three-sublattice color-ordered phase (the analogue of the N\'eel phase) and a trimerized $SU(3)$ singlet phase (the analogue of the VBS phase). This theory consists of three $CP^2$-models coupled by quartic interaction terms. Employing the functional renormalization group method in a suitable large $N$ limit, we located a stable critical fixed point, which is not in the universality class of the $CP^n$-model. 

Our derivation of the critical field theory starts from the magnetically ordered phase. One drawback of this approach is that the properties of the paramagnetic state are encoded in subtle Berry phase effects, which we did not analyze in this work. However, we provide a strong argument that our field theory correctly describes the deconfined critical point by presenting an alternative derivation of the same critical field theory starting from the paramagnetic VBS phase (details can be found in App.~\ref{VBStheorysec}). Nevertheless, a detailed analysis of Berry phase effects in the spirit of Haldane's work \cite{Haldane1988}, checking that 
they result in a sextupling of hedgehog events, implying the non-compactness of our critical theory and eventually giving rise to the six-fold degenerate trimer phase, would further substantiate our claim and we leave this problem open for future study.

Additionally, in order to substantiate our understanding of the critical action, it would be interesting to analyze the constraint-softening that leads to its final polynomial form. One possible way to achieve this would be to combine the three mutually orthogonal spinors to an $SU(3)$ matrix order parameter. The resulting theory is then a matrix field theory where the orthogonality constraint is rigorously incorporated (see also Ref.~\cite{smerald2013theory}). 

A more advanced RG study of our critical field theory would be worthwhile as well. Even though our fRG analysis shows that a new stable critical fixed point exists, the fact that it appears only at rather large $N$ is clearly a shortcoming of our simple truncation scheme. Similar problems are well known in the context of the $CP^n$ model, where simple RG approaches give rise to runaway RG flows at small $N$. Nevertheless, we expect that the large $N$ critical fixed point survives in the limit $N \to 2$. Within fRG this could be analyzed using a more sophisticated truncation scheme.

Finally, one can easily generalize our approach to other interesting $SU(N)$ magnets in two-dimensions, such as $SU(4)$ on the square lattice, where a dQCP between a magnetically ordered phase and a quadrimerized singlet phase is possible.

\acknowledgements
The authors acknowledge insightful discussions with Sebastian Huber, Dennis Schimmel and T. Senthil. This work was supported by the German Excellence Initiative via the Nanosystems Initiative Munich (NIM).

\appendix

\titleformat{\subsection}
  {\normalfont\itshape}{\thesubsection}{1em}{\centering}

\titleformat{\section}
			{\normalfont\bfseries}{Appendix \thesection:\!}{.5em}{\normalfont\bfseries}

\section{Derivation of the flow equations}	
\label{FRGapp}		
						
In this appendix, we present the derivation of the flow equations (\ref{mfloweq})--(\ref{etafloweq}). Let us first 
explicitly denote the gauge fixing term by 
\begin{align}
\Gamma_\text{gf} = \frac{1}{2\alpha_{\text{gf}}} \sum_i\int d^D x (\partial_\mu A^i_\mu)^2  \ .
\end{align}
We will work in the limit $\alpha_{\text{gf}} \rightarrow 0$, which fixes the transversal gauge.

Next, we specify the regularization procedure. Following \cite{bartosch2013corrections}, we employ the Litim regulator $\mathcal{R}_\Lambda[\A]$, which will allow for simple analytic evaluations since momentum integrals are rendered trivial. When evaluated at zero gauge field, its scalar and gauge field components in momentum space take the form 
\begin{align}
\label{Litim regulator}
&R_\Lambda^{\phi}(\k) = Z_\Lambda (\Lambda^2 - k^2) \theta(\Lambda^2 - k^2) \\ 
& R_\Lambda^A(\k) = \frac{1}{e^2_\Lambda}\left(\Lambda^2 - k^2\right) \theta(\Lambda^2 - k^2)\ .
\end{align}
At nonzero gauge fields, we need to replace ordinary derivatives by covariant ones in the real-space version of the scalar regulator (\ref{Litim regulator}), but this will only be of relevance for the flow of gauge coupling, to be discussed therein.

Having specified all ingredients to the Wetterich equation (\ref{Wetterich}) (except for the counterterm $\mathcal{C}_\Lambda$, see below), we can compute the running of the couplings. The running of the mass term reads
\begin{align}
\label{massrunningveryfirst}
\frac{\partial}{\partial l} m_\Lambda = \frac{1}{2} \text{Tr} \left[ \frac{\partial}{\partial l} \mathcal{R}_\Lambda \left( - \hat{G} \frac{\delta^2 \hat{\Gamma}_\Lambda^{(2)}}{\delta \overline{\phi}_{11}(\0) \delta \phi_{11}(\textbf{0})} \hat{G} \right) \right]  \bigg|_{\phi=0, \A=0} \ ,
\end{align}
where the argument of $\phi_{11}, \bar{\phi}_{11}$ denotes zero momentum.
The propagator matrix $\hat{G}$ has the following structure in field derivative space 
\begin{align}
\hat{G} =  
\bordermatrix{
  & \delta\phi	 & \delta\bar{\phi}   & \delta A  \cr
\delta\phi& 0 & G_\phi & 0  \cr
\delta\bar{\phi} & G_\phi & 0 & 0  \cr
\delta A & 0 & 0 & G_A  \cr
} \ , 
\end{align}
with scalar and gauge field components in the transversal gauge 
\begin{align}
\label{propagatorphi}
&G_\phi(\p, \beta, j; \k, \alpha, i) = \delta_{ij} \delta_{\alpha\beta}\delta(\k - \p) \cdot \frac{1}{Z_\Lambda k^2 + m_\Lambda + R_\Lambda^\phi(k)} \\ 
\label{propagatorA}
& G_A(\p,\nu,j; \q, \mu,i) =
 \delta_{ij} \delta(\q + \p)\cdot \frac{\delta_{\mu\nu} - q_\mu q_\nu/q^2}{q^2/e_\Lambda^2 + R_\Lambda^A(q)} \ . 
\end{align}
The same field space structure applies to the regulator matrix $\mathcal{R}_\Lambda$, with components as given in (\ref{Litim regulator}). Performing the trace over field space in (\ref{massrunningveryfirst}) results in  
\begin{align}
\nonumber
 & \frac{\partial}{\partial l} m= \text{tr}\! \intk\bigg[ \frac{\partial}{\partial l} R_\Lambda^{\phi}(\k) \left(\! - G_{\phi}(\k) \frac{\delta^2 \Gamma_{\bar{\phi}\phi}(\k)}{\delta \overline{\phi}_{11}(\0) \delta \phi_{11}(\0)} G_{\phi}(\k) \! \right)   + \\ &\frac{1}{2}\frac{\partial}{\partial l} R_\Lambda^{A}(\k) \left( \!- G_{A}(-\k) \frac{\delta^2 \Gamma_{AA}(\k)}{\delta \overline{\phi}_{11}(\0) \delta \phi_{11}(\0)} G_{A}(-\k) \right) \bigg]  \bigg|_{\substack{\phi=0 \\ \A=0}} \ ,
\label{massflowfirst}
\end{align}
where $\text{tr}$ denotes the sum over all discrete indices, 
\begin{align*}
\intk \equiv \int \frac{d^3k}{(2\pi)^3} \ ,
\end{align*}
and $\Gamma_{\bar{\phi}\phi}, \Gamma_{AA}$ are the field space components of the vertex operator $\hat{\Gamma}_\Lambda^{(2)}$ in obvious notation.

It should be noted that diagrams for the mass flow involving 3-point vertices are absent in the transversal gauge.
The running of the quartic couplings $\rho^\Lambda_{1,2,3}$ can be obtained in analogous manner, using 
\begin{align}
\nonumber
 &\partial_l \rho_1^\Lambda = \frac{1}{4}\frac{\delta^4  \partial_l\Gamma_\Lambda}{\delta \overline{\phi}_{11}(\0)\delta\phi_{11}(\0)\delta\overline{\phi}_{11}(\0) \delta{\phi}_{11}(\0)}\bigg|_{\phi=0, \A=0} \\  
 \nonumber &\partial_l \rho_2^\Lambda = \frac{1}{2}\frac{\delta^4 \partial_l \Gamma_\Lambda}{\delta \overline{\phi}_{11}(\0)\delta\phi_{11}(\0)\delta\overline{\phi}_{22}(\0) \delta{\phi}_{22}(\0)}\bigg|_{\phi=0, \A=0} \\  \quad &\partial_l \rho_3^\Lambda = \frac{1}{2}\frac{\delta^4 \partial_l \Gamma_\Lambda}{\delta \overline{\phi}_{11}(\0)\delta\phi_{12}(\0)\delta\overline{\phi}_{22}(\0) \delta{\phi}_{21}(\0)}\bigg|_{\phi=0, \A=0}
\end{align}
For general external indices the required fourfold derivative reads 
\begin{align}
\label{quartic full 1} \nonumber
&\frac{\delta^4  \partial_l\Gamma_\Lambda}{\delta \overline{\phi}_{\alpha_{4}i_{4}}(\0)\delta\phi_{\alpha_{3}i_{3}}(\0)\delta\overline{\phi}_{\alpha_{2}i_{2}}(\0) \delta{\phi}_{\alpha_{1}i_{1}}(\0)}\bigg|_{\phi=0, \A=0} = \\ \notag &\frac{1}{2}\text{Tr}\left[\frac{\partial}{\partial l}\mathcal{R}_\Lambda \left(  \hat{G} \frac{\delta^2 \hat{\Gamma}_\Lambda^{(2)}}{\delta \overline{\phi}_{\alpha_{4} i_{4}}(\0)\delta \phi_{\alpha_{3} i_{3}}(\0)}\hat{G} \frac{\delta^2 \hat{\Gamma}_\Lambda^{(2)}}{\delta \overline{\phi}_{\alpha_{2} i_{2}}(\0)\delta \phi_{\alpha_{1} i_{1}}(\0)}\hat{G}    
\right.\right.  \\ & \quad + \text{permutations} \bigg) \bigg] \ ,
\end{align}
where all possible permutations of the external field derivatives acting on $\hat{\Gamma}_\Lambda^{(2)}$ need to be taken into account.
Performing the trace over field space yields, for the flow of $\rho_1^\Lambda$, a similar structure as Eq.~(\ref{massflowfirst}), see Fig.\ \ref{diagrampics}. By contrast, the flow equations for $\rho_{2/3}^\Lambda$ do not pick up any contributions from the gauge sector, which is a consequence of the lattice dependence of the gauge fields. 

Similarly, the running of the wave-function renormalization can be extracted from 
\begin{widetext}
\begin{align}
 \notag
& \frac{\partial}{\partial l} Z_\Lambda = 
\frac{\partial}{\partial p^2} \frac{\delta^2 \partial_l \Gamma_\Lambda}{\delta \overline{\phi}_{11}(\p)\delta {\phi}_{11}(\p)} \bigg|_{p=0} = 
\frac{\partial}{\partial p^2} \frac{1}{2} \text{Tr} \left[\frac{\partial}{\partial l} \mathcal{R}_\Lambda \left(\hat{G} \frac{\delta \hat{\Gamma}_\Lambda^{(2)}}{\delta \overline{\phi}_{11}(\p)} \hat{G} \frac{\delta \hat{\Gamma}_\Lambda^{(2)}}{\delta {\phi}_{11}(\p)} \hat{G} 
+ \phi_{11} \leftrightarrow \bar{\phi}_{11} \right)\right]\bigg|_{p=0} =  \\ & 
\frac{\partial}{\partial p^2} \frac{1}{2} \text{tr}\intk \left[G_{A}(\k) \left[\frac{\delta \Gamma_{A \phi}}{\delta \overline{\phi}_{11}(\p)}  \right]\!(-\k, \p + \k ,\p) \  G_{\phi}(\p + \k)\left[\frac{\delta \Gamma_{\overline{\phi} A}}{\delta {\phi}_{11}(\p)}  \right] \!(\k, \p, \p + \k) \ G_A(\k) \partial_l R_\Lambda^A(-\k)  + \phi_{11} \leftrightarrow \bar{\phi}_{11}\right] \bigg|_{p=0} \ , \label{Zl flow first}
\end{align}
\end{widetext}
where the second term in (\ref{Zl flow first}), obtained by permuting the fields, has a similar momentum structure as the first one. 

To evaluate the flow equations (\ref{massflowfirst}), (\ref{quartic full 1}), (\ref{Zl flow first}), one needs to insert the appropriate vertex terms. They read 
\begin{align} \notag
&\frac{\delta^{4} \Gamma_\Lambda}{\delta \overline{\phi}_{\alpha_{4}i_{4}}(\0)\delta\phi_{\alpha_{3}i_{3}}(\0)\delta\overline{\phi}_{\alpha_{2}i_{2}}(\k) \delta{\phi}_{\alpha_{1}i_{1}}(\p)} = \\  & \left(
2\hat{\rho}^\Lambda_{i_1 i_2 i_3 i_4} \delta_{\alpha_1 \alpha_2} \delta_{\alpha_3 \alpha_4} + 2\hat{\rho}^\Lambda_{i_1 i_4 i_3 i_2} \delta_{\alpha_2 \alpha_3} \delta_{\alpha_1 \alpha_4} \right) \delta(\k - \p)
\\
&
\frac{\delta^{4} \Gamma_\Lambda}{\delta \overline{\phi}_{11}(\0) \delta \phi_{11}(\0) \delta A^1_\mu(\k) \delta A^1_\nu(\p)} = 2Z_\Lambda \delta_{\mu\nu} \delta(\k + \p) 
\\
& \notag
\frac{\delta^{3} \Gamma_\Lambda}{\delta \overline{\phi}_{11}(\p) \delta \phi_{\alpha i}(\q) \delta A^j_\mu(\k)} =
\\ & -Z_\Lambda \delta_{i1}\delta_{j1}\delta_{\alpha 1} \delta(\p - \q - \k) \cdot \left(p_\mu + q_\mu\right) \ . 
\end{align}
Inserting these vertices, and the Litim cutoff, all momentum integrations are rendered trivial, and the flow equations are readily computed -- see main text, Eqs.~(\ref{mfloweq}) -- (\ref{etafloweq}).

The flow of the inverse gauge coupling $(1/e^2_\Lambda)$ could be derived in a similar manner, by taking an appropriate momentum derivative of the diagram shown in Fig. \ref{diagrampics}.
However, to avoid ambiguities arising from the sharp Litim cutoff, we instead follow the recipe presented by Reuter and Wetterich in \cite{reuter1993running, reuter1994exact}. Adapted to our lattice-dependent gauge field setup, its main idea is as follows: 
We start from $\Gamma_\Lambda$ in real space and evaluate it at a field configuration where $\phi, \bar{\phi}, \A^2,\A^3 =0$ and the gauge field of sublattice $1$ is such that it corresponds to a constant magnetic field $B$. Then 
\begin{align}
\Gamma_\Lambda = \int d^3 x \frac{1}{4 e^2_\Lambda} \cdot 2B^2 = \frac{1}{e^2_\Lambda} \cdot \frac{1}{2} B^2 \Omega \quad  , 
\end{align}
where $\Omega$ is the system volume. Now, the flow of $1/e^2_\Lambda$ can be obtained by evaluating the right hand side of the Wetterich equation (\ref{Wetterich}) in the field configuration described above, and singling out the coefficent proportional to $B^2\Omega$. By construction of the effective action, on the r.h.s. of (\ref{Wetterich}) the vector potential $\A$ enters only via the squared covariant derivative 
$-D\left(\textbf{A}\right)^2$, where $D(\textbf{A})_\mu = \partial_\mu - iA_\mu$. Fortunately, the spectrum of $-D(\A)^2$ in the given field configuration is explicitly known; it is related to a summation over Landau levels. Using the Euler-McLaurin summation formula to evaluate the sum, the flow of the gauge coupling is then derived as 
\begin{widetext}
\begin{align}
\label{bypartsfirst}
\frac{\partial}{\partial l} \left(\frac{1}{e^2_\Lambda}\right) = \frac{1}{24\pi^2}  \sum_{\alpha = 1}^{N} \int_{0}^\infty dx\  x^{-1/2} \frac{d}{dx}\left(\partial_l R_\Lambda^\phi(x) \frac{1}{Z_\Lambda x+m_\Lambda+R_\Lambda^\phi(x)}\right) +  \frac{\partial}{\partial l} \left(\frac{1}{e^2_\Lambda}\right)   \bigg |_{\mathcal{C}_\Lambda} \ , 
\end{align}
\end{widetext}
where the second summand denotes the contribution from $\mathcal{C}_\Lambda$ specified below. 
To compute the integral, it is convenient to use the prescription
\begin{align}
\label{epsilontik1}
\lim_{\epsilon \rightarrow 0} \lim_{\delta \searrow 0} \frac{N}{24\pi^2} \int_{\delta}^\infty dx\  x^{-1/2} \frac{d}{dx}\left( \frac{\partial_l R_\epsilon(x)}{Z_\Lambda x+m_\Lambda+R_\epsilon(x)}\right) \ , 
\end{align}
where $R_\epsilon$ is a version of the Litim cutoff smoothened over a small momentum range $\epsilon$. Then, integrating by parts, the integral is easily solved, and the limits can be taken without problems.  

Let's now consider the extra term $\mathcal{C}_\Lambda$. 
Following \cite{reuter1994exact}, an appropriate form of $\mathcal{C}_\Lambda$, which cancels the spurious background-field derivatives from the Wetterich equation, is given by
\begin{align}
\label{Cansatz}
\mathcal{C}_\Lambda [\A] \simeq -\frac{1}{2} \text{Tr}_\phi \left [ \log\left(1+ \frac{R_\Lambda^\phi[\A]}{Z_\Lambda \Lambda^2 + m}\right)\right ] \  .
\end{align}
The notation $\text{Tr}_\phi$ indicates that the trace only extends over the scalar sector in field space; in general, $\mathcal{C}_\Lambda$ also has contributions from the gauge sector, but these are not of relevance for the flow equations at our level of truncation.

One should note that the ansatz Eq.~\eqref{Cansatz} works best for regulators whose $l$-derivatives are strongly peaked near momenta $\simeq \Lambda$. This does not hold true for the Litim regulator. However, as already mentioned, the error made this way is of two-loop order. To be on the safe side, we have also checked that a Litim-adaption of Eq.~\eqref{Cansatz} with an extra parameter does not modify the qualitative analytical structure of the stable fixed point we are after.

To compute the $\mathcal{C}_\Lambda$-contribution to the running of $1/e^2_\Lambda$, following the logic above we need to evaluate the $l$ derivative of (\ref{Cansatz}) in the field configuration corresponding to the constant magnetic field on sublattice 1. Again using the Landau-Level summation, this leads to 
\begin{align}
\notag
&\frac{\partial}{\partial l} \left(\frac{1}{e^2_\Lambda}\right)\bigg|_{\mathcal{C}_\Lambda}  = \\  &- \frac{N}{24 \pi^2} \frac{\partial}{\partial l} \int_0^\infty dx x^{-1/2} \frac{\partial}{\partial x}  \log  \left(1 + \frac{R_\Lambda^\phi(x)}{Z_\Lambda\Lambda^2 + m_\Lambda}\right) .
\end{align}
This form can easily be evaluated analytically.
Rephrasing the so-derived flow equations in terms of dimensionless quantities, one obtains the flow equations as given in the main text, Eqs.\ (\ref{mfloweq}) -- (\ref{etafloweq}).

\section{Field theory from the trimer side}
\label{VBStheorysec}

In this appendix we present an alternative construction of the critical theory in Eq.~\eqref{finalaction} by starting from the trimerized VBS phase.
To this end, let us first recapitulate the approach by Levin and Senthil for the the $SU(2)$ case on the square lattice \cite{levin2004deconfined}, which we will follow here.

\begin{figure}
\centering
\includegraphics[width=\columnwidth]{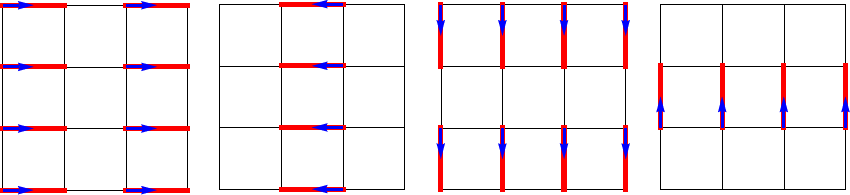}
\caption{Sketch of the four degenerate groundstates of the $SU(2)$ VBS. The red lines indicate $SU(2)$-singlets,  the blue arrows represent the $\mathbb{Z}_4$-order. This picture was adapted from \cite{levin2004deconfined}.}
\label{SU2VBS}
\end{figure} 

\begin{figure}
\centering
\includegraphics[width=0.8 \columnwidth]{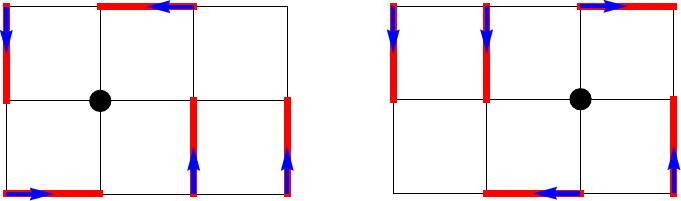}
\caption{Sketch of the vortex and antivortex in the $SU(2)$ case. The black circle indicates the vortex core, which carries a free spin. A translation by a lattice unit vector takes a vortex into an antivortex.}
\label{SU2vortices}
\end{figure} 

The columnar $SU(2)$ VBS order can be described by a $\mathbb{Z}_4$ clock order parameter and admits four degenerate ground states, sketched in Fig.~\ref{SU2VBS}.   The natural field theory that captures this $\mathbb{Z}_4$ order is an XY-model with a quartic anisotropy term $\cos(4\theta)$, which is known to be dual to a compact $U(1)$ gauge theory with magnetic monopoles by means of the particle vortex duality \cite{PhysRevLett.47.1556}.
The topological excitations of this model are $\mathbb{Z}_4$-vortices, where four VBS domain walls merge at a single point. On the gauge theory side, they correspond to electrically charged bosons. Proliferating these vortices destroys the VBS order, consequently they are a natural starting point to construct a theory for the deconfined QCP. Levin and Senthil now made the following crucial observations: i) the vortices always carry a free spin at the core (see the sketch in Fig.\ \ref{SU2vortices}). Therefore, the corresponding field theory is a modified XY-model with an additional spinor structure, or equivalently a $U(1)$ gauge theory of charged bosons $z_\alpha$ with charge $q=1$, which transform as spinors under $SU(2)$
\begin{equation}
z_{\alpha} \  \overset{SU(2)}{\longrightarrow} \ U_{\alpha \beta} z_{\beta }  \hspace{1cm} z_{\alpha } \ \overset{U(1)}{\longrightarrow} \ e^{i \phi} z_{\alpha } \ . 
\end{equation}
ii) the $\mathbb{Z}_4$-structure of the vortices is encompassed in the $\cos(4\theta)$-term. For the standard XY-model, this term is known to be irrelevant at the quantum critical point, and it is reasonable to expect that this will be the case for the modified XY-model as well. 
iii) expanding the action in the spinor fields $z_\alpha$ and their derivatives, one therefore immediately arrives at the standard $CP^1$-model in Eq.~\eqref{cp1model}.

\begin{figure}
\centering
\includegraphics[width=\columnwidth]{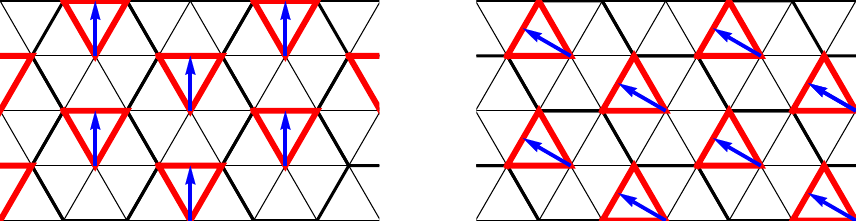}
\caption{Two out of six degenerate ground states of the trimer state. The red triangles indicate $SU(3)$-singlets, the blue arrows represent the $\mathbb{Z}_6$ order.}
\label{VBStwoex}
\end{figure}

The Higgs-phase of the $CP^1$-model can then be identified with the N\'eel state as follows. Vortices are always centered on one sublattice, while antivortices are centered on the other sublattice. A lattice translation by one unit vector thus takes a vortex into an antivortex (see Fig. \ref{SU2vortices}). Both vortices and antivortices must transform under the fundamental representation of $SU(2)$, since both carry a free spin. Furthermore, the antivortex must have $U(1)$ gauge charge $q=-1$. As the vortex is represented by $z_\alpha$, the antivortex must be represented by $\epsilon_{\alpha\beta}\bar{z}_\beta$, which fullfills both these requirements. Therefore, a translation operation acts as 
\begin{align}
\label{spinorshift}
z_\alpha \rightarrow \epsilon_{\alpha\beta} \bar{z}_\beta  \ ,
\end{align}
flipping the vortex spin. This immediately shows that condensation of vortices results in antiferromagnetic N\'eel order.

The $CP^1$ model in Eq.~(\ref{cp1model}), obtained as critical continuum theory from this elegant reasoning, does not make any explicit reference to the sublattice structure. This structure is hidden in the transformation property of the spinors (\ref{spinorshift}). Note that a lattice translation is equivalent to time reversal in this case (which makes sense since it turns a vortex into an antivortex), and so also flips the spatial components of the gauge field $\mathcal{A}_j \rightarrow -\mathcal{A}_j$.

Let's now make the sublattice structure explicit by assigning the field $z_{\alpha1}$ to the vortex (which sits on sublattice 1) and the field $z_{\alpha2}$ to the antivortex (which sits on sublattice 2). As a lattice translation takes $z_{\alpha1}$ to $z_{\alpha2}$, Eq.\ (\ref{spinorshift}) translates to 
\begin{align}
\label{SU2fixedconstraint}
z_{\alpha 2} = \epsilon_{\alpha\beta}\bar{z}_{\beta 1} \ .
\end{align}
This relation can also be understood from the simple reasoning that it must be possible to break up a $SU(2)$ singlet in the VBS phase into a vortex - antivortex pair or vice versa. In the Lagrangian this process corresponds to a term of the form 
\begin{equation}
\varepsilon_{\alpha \beta} z_{\alpha1} z_{\beta 2} \equiv \bar{z}_{\alpha 1} z_{\alpha 1}
\end{equation}
which is indeed an $SU(2)$ singlet and charge neutral under $U(1)$, as required.

When we derived the $SU(3)$ critical theory from the $3$-color-ordered side in Sec.~\ref{Pathintsec}, we asserted that a $SU(2)$ derivation along the same lines was clearly possible; this derivation  results in Eq.~(\ref{SU2propconstraint}): $z_{\alpha 2} \sim \epsilon_{\alpha\beta}\bar{z}_{\beta 1}$. That is, one obtains the same relation as in Eq.~(\ref{SU2fixedconstraint}), but without fixed relative phase. The fact that the relative phase is not fixed will a priori lead to two different, sublattice dependent gauge fields $\mathcal{A}_\mu^i$. However, this enlarged gauge redundancy does not lead to different properties of the dQCP in the $SU(2)$ case, because the two $U(1)$ gauge fields couple to the field $z_\alpha$ in precisely the same way after using Eq.~\eqref{SU2propconstraint}.

To summarize, for the $SU(2)$ case the difference between our derivation and the VBS derivation by Levin and Senthil lies solely in the fact that the relative phase of the fields $z_{\alpha i}$ is fixed by the latter. 
 
Let us now attempt a derivation of our critical theory for the $SU(3)$ case from the trimer side.
We will only consider coverings of the triangular lattice with $SU(3)$ singlets which have the same unit cell as the $3$-color-ordered state (3 times the unit cell of the triangular lattice). There are six such coverings, corresponding to six degenerate paramagnetic ground states. Two of them are shown in Fig.\ \ref{VBStwoex}.
Consequently a natural description of this trimer state is a $\mathbb{Z}_6$ clock model, respectively an XY model with sixfold anistropy $\cos(6\theta)$. Again, the anisotropy term is irrelevant at the critical point, so we're not going to consider it further.

\begin{figure}
\centering
\includegraphics[width=.6\linewidth]{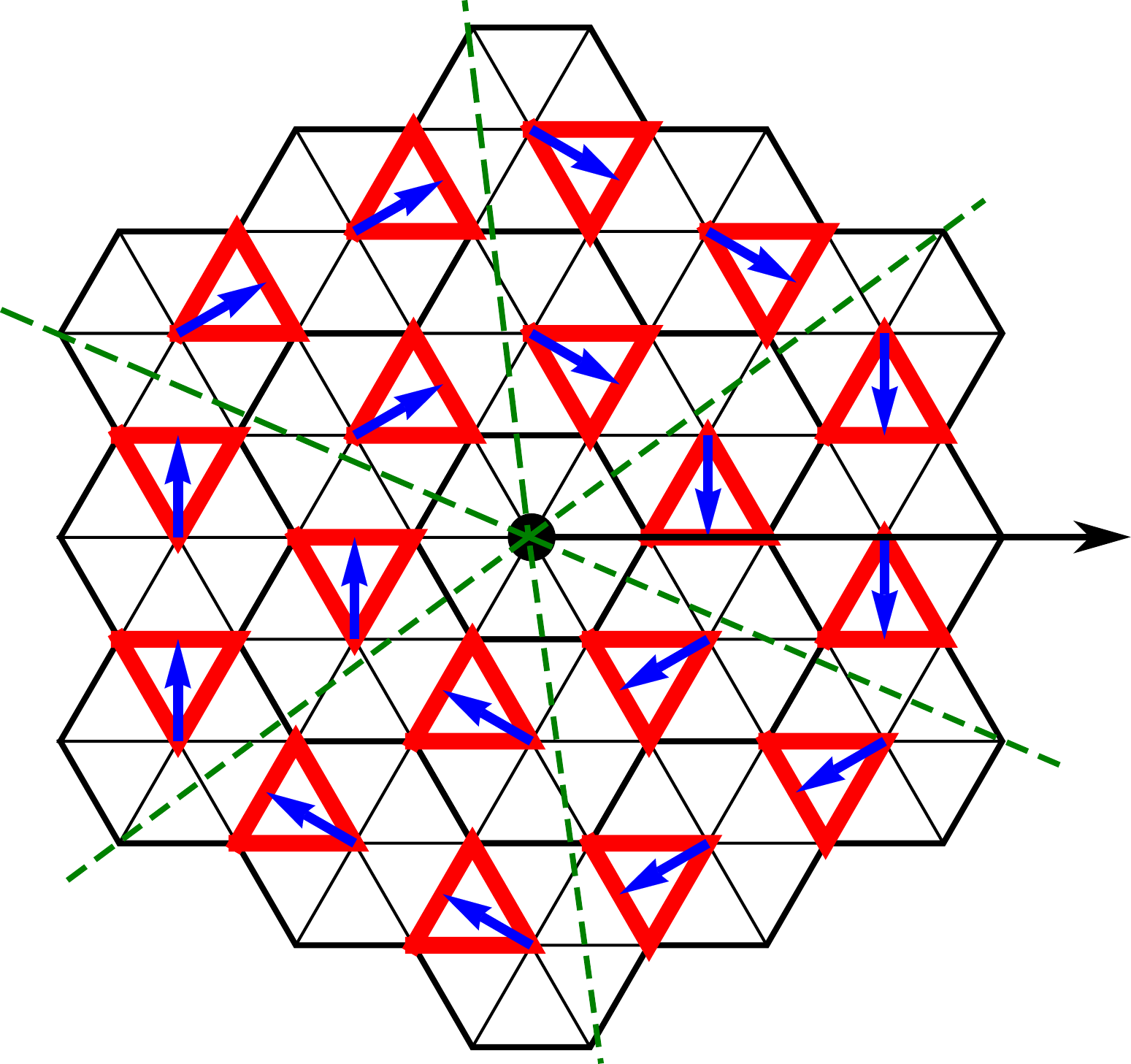}
\caption{A vortex $z_{\alpha 1}$ centered on sublattice 1. The black circle represents the free $SU(3)$ spin and the dashed green lines indicate the domain walls.}
\label{firstvortex}
\end{figure}

In analogy to the $SU(2)$ case it is easy to see that $\mathbb{Z}_6$ vortices carry a free $SU(3)$ spin, as shown in Fig.~\ref{firstvortex}. Interestingly, it is not possible to draw a corresponding antivortex, however. As we will argue below, this is due to the fact that an antivortex transforms under the conjugate representation in the $SU(3)$ case and thus cannot carry a single free $SU(3)$ spin at its core.

The fundamental degrees of freedom in our theory are thus bosonic fields $z_{\alpha j}$ which describe a vortex on sublattice $j$ transforming under the fundamental representation of $SU(3)$, coupled to a $U(1)$ gauge field
\begin{equation}
z_{\alpha j} \  \overset{SU(3)}{\longrightarrow} \ U_{\alpha \beta} z_{\beta j}  \hspace{1cm} z_{\alpha j} \ \overset{U(1)}{\longrightarrow} \ e^{i \phi_j} z_{\alpha j} \ .
\end{equation}
As a next step, we need to find an analogue of Eq.~\eqref{SU2fixedconstraint} for the $SU(3)$ case in order to determine how antivortices are represented in our theory. By the above rationale we should aim at a theory with a fixed relative phase. As we argue below, the direct generalization of (\ref{SU2fixedconstraint}) to our case of interest is given by (summation convention over repeated indices is implied)
\begin{align}
\label{SU3fixedconstraint}
\bar{z}_{\alpha i} = \epsilon_{ijk}\epsilon_{\alpha\beta\gamma} z_{\beta j} z_{\gamma k} \ , 
\end{align} 
which only works if we simultaneously demand $|z_{\alpha i}|=1$ for all three sublattices $i$. Again, this equation can be understood from the fact that it must be possible to break up a $SU(3)$ singlet in the VBS into three vortices on the respective neighboring sublattice sites or vice versa. The corresponding term in the Lagrangian is then
\begin{equation}
\varepsilon_{ijk} \varepsilon_{\alpha \beta \gamma} z_{\alpha i}  z_{\beta j}  z_{\gamma k} \equiv \bar{z}_{\alpha i} z_{\alpha i} \ ,
\end{equation}
which is clearly an $SU(3)$ singlet, as required. More importantly, this $SU(3)$ singlet also needs to be charge neutral under $U(1)$ transformations. Obviously, this is only possible if we allow for sublattice dependent $U(1)$ gauge transformations. Eq.~\eqref{SU3fixedconstraint} thus enforces a partial gauge fixing condition $\phi_1+\phi_2+\phi_3=0$.
Note that Eq.~\eqref{SU3fixedconstraint} ensures that the spinors on the three sublattices are mutually orthogonal. The condensation of vortices thus automatically leads to a three-sublattice color ordered phase, as discussed in the main text. Moreover, Eq.~\eqref{SU3fixedconstraint} is in accordance with $SU(3)$ transformation properties of the spinors, i.e. $\bar{z}_{\alpha j}$ transforms under the conjugate representation of $SU(3)$.

\begin{figure}[H]
\centering
\includegraphics[width=\columnwidth]{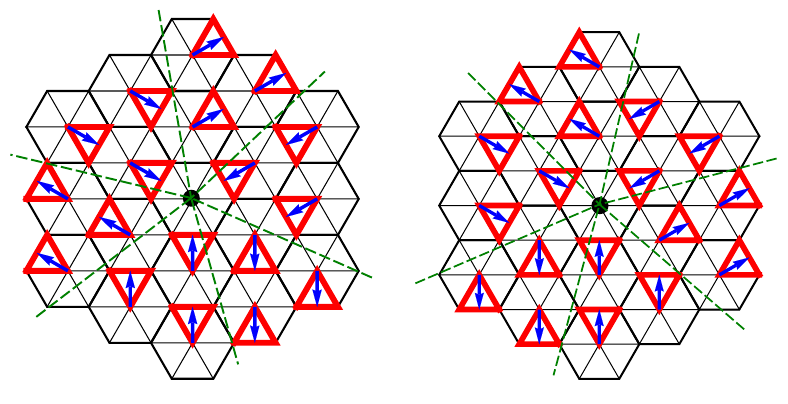}
\caption{Two composite vortices centered on sublattice 1, which we represent by the fields $x_{\alpha 1}$ (left) and $y_{\alpha 1}$ (right), respectively. We identify vortices which are obtained from $x_{\alpha i},y_{\alpha i}$ by space-reflection along the axis indicated in Fig.~\ref{firstvortex} with $x_{\alpha i},y_{\alpha i}$, accordingly. }
\label{3vortices}
\end{figure}

In the following we present a more microscopic justification of Eq.~\eqref{SU3fixedconstraint} by deriving it from the lattice transformation properties
of all topological excitations which can be represented pictorially. In addition to the vortex of Fig.~\ref{firstvortex}, we can draw two ``composite vortices'' on each sublattice, which we denote by $x_{\alpha i}, y_{\alpha i}$, respectively. An example is depicted in Fig.~\ref{3vortices}. Due to the discrete nature of the $\mathbb{Z}_6$ order parameter, one cannot assign a unique vorticity (or gauge charge) $q$ to these objects. The most natural choice seems to be $q=-2$, but this assignment is ambiguous as e.g.\ the vorticity is not reversed if the vortex is encircled clockwise.

\begin{figure}
\centering
\includegraphics[width=.5\linewidth]{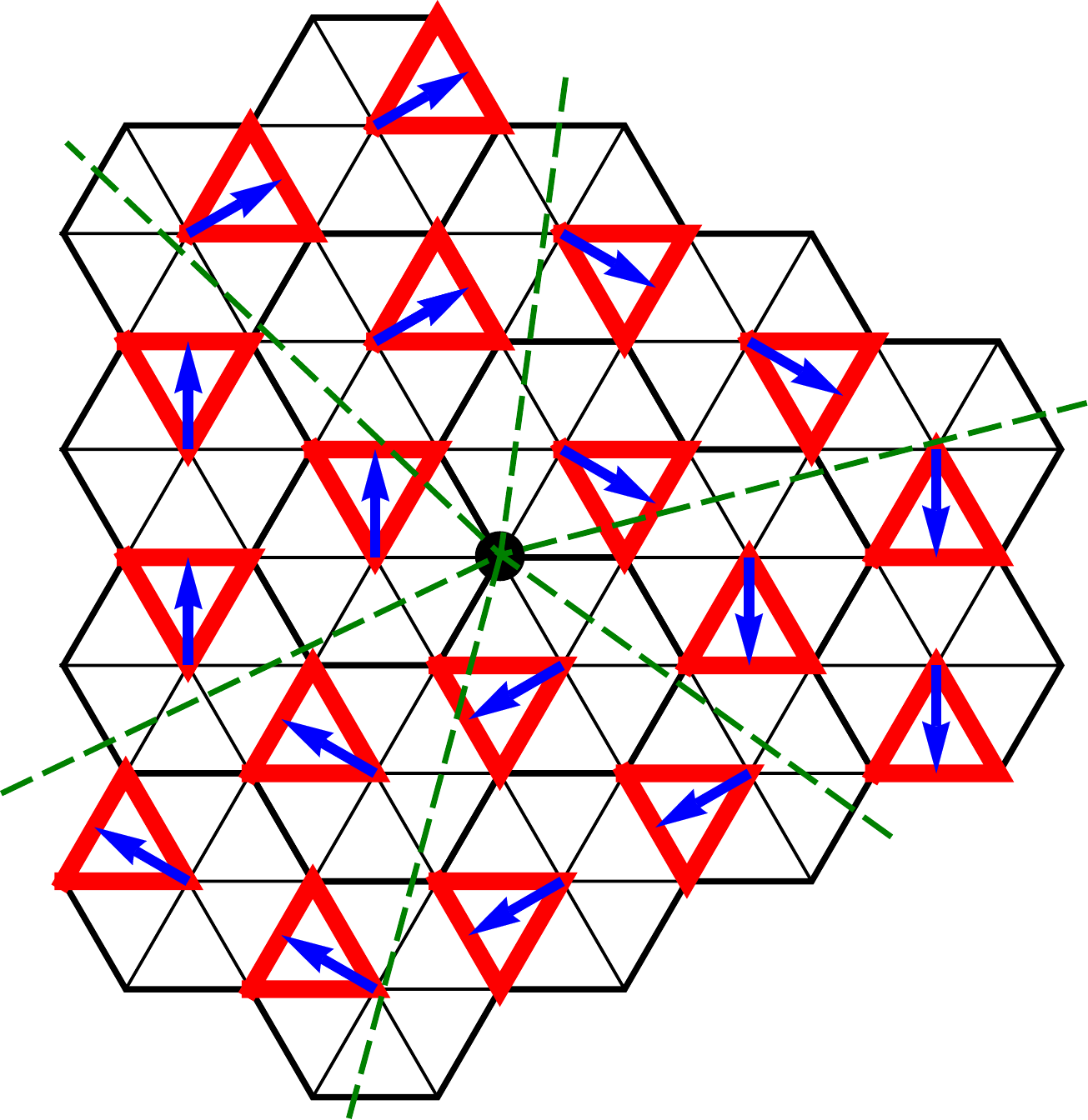}
\caption{A $z_{\alpha 2}$-vortex centered on sublattice 2, obtained by a shift of a vortex $y_{\alpha 1}$ on sublattice 1 (see Fig. \ref{3vortices}) to the right by one unit vector. }
\label{fig:vortex_shifted}
\end{figure}

 Analogous pictures can be drawn for vortices centered on sublattices $2,3$, which can be obtained from Fig. \ref{3vortices} by translation operations. One such operation is sketched in Fig.\ \ref{fig:vortex_shifted}.
A highly interesting feature of these pictures is that, opposed to the $SU(2)$-case, one can never draw an antivortex as an elementary object carrying a free $SU(3)$ spin. 

Let us now try to parametrize the composite vortices $x,y$ in terms of the elementary one. Up to irrelevant constant prefactors, the only meaningful ansatz is
\begin{align}
\label{doublevortex}
x_{\alpha i} = y_{\alpha i} = \epsilon_{ijk} \epsilon_{\alpha\beta\gamma} \bar{z}_{\beta j} \bar{z}_{\gamma k} \ , 
\end{align}
which is consistent with transformations under the fundamental representation of $SU(3)$. 

Next, we need to take into account the transformation properties of the vortices. Drawing all possible pictures, they read 
\begin{align}
\label{Rtrans}
&\hat{R} \ (\text{Rot.\ by $\pi/3$ with base point on sublattice 1}):  \\ &z_1 \leftrightarrow z_1, x_1 \notag  \leftrightarrow y_1, z_2 \leftrightarrow z_3, x_2 \leftrightarrow y_3 , y_2 \leftrightarrow x_3  \\
\notag & \\
\label{T1trans}
&\mathcal{T}_1 (\text{transl.\ along the first lattice direction}): \\ \notag
&z_1 \rightarrow x_2 \rightarrow y_3 \rightarrow z_1 \\
\notag 
&x_1 \rightarrow y_2 \rightarrow z_3 \rightarrow x_1 \\
\notag 
&y_1 \rightarrow z_2 \rightarrow x_3 \rightarrow y_1 \\
 \notag & \\
\label{T2trans}
&\mathcal{T}_2 (\text{transl.\ along the second lattice direction}): \\ \notag
&z_1 \rightarrow y_3 \rightarrow x_2 \rightarrow z_1 \\
\notag 
&x_1 \rightarrow z_3 \rightarrow y_2 \rightarrow x_1 \\
\notag 
&y_1 \rightarrow x_3 \rightarrow z_2 \rightarrow y_1 
\end{align}
Here, we have supressed the $SU(3)$-indices. Along with reflection along the $x$-axis, which acts trivially on the vortices per definition (see caption of Figs.\ \ref{firstvortex}, \ref{3vortices}), the operations $\hat{R}, \mathcal{T}_{1/2}$ span the symmetry group of the triangular lattice. 
Note that the above transformations hold up to constant prefactors, which cannot be deduced from the pictures. One must now reconcile the transformation properties (\ref{Rtrans}-\ref{T2trans}) with the definition of the composite vortices (\ref{doublevortex}). It is found that the only way to do so is precisely to demand the orthogonality constraint with fixed relative phase (\ref{SU3fixedconstraint})! Writing down a gauge theory consistent with (\ref{SU3fixedconstraint}), one then recovers Eq.~(\ref{intermediateZ}) along with the relative gauge fixing constraint. Again, softening the unit length constraint for the fields $z_{\alpha j}$ we arrive at the theory in Eq.~\eqref{finalaction}.

\bibliographystyle{apsrev4-1}

\bibliography{su3dcp}

\end{document}